\documentclass[12pt,preprint]{aastex}

\newcommand{\noprint}[1]{}
\newcommand{\figsetstart}{{\bf Fig. Set} }
\newcommand{\figsetend}{}
\newcommand{\figsetgrpstart}{}
\newcommand{\figsetgrpend}{}
\newcommand{\figsetnum}[1]{{\bf #1.}}
\newcommand{\figsettitle}[1]{ {\bf #1} }
\newcommand{\figsetgrpnum}[1]{\noprint{#1}}
\newcommand{\figsetgrptitle}[1]{\noprint{#1}}
\newcommand{\figsetplot}[1]{\noprint{#1}}
\newcommand{\figsetgrpnote}[1]{\noprint{#1}}

\def\cm#1{\ifmmode {\,{\rm cm^{-#1}}}                  
        \else \hbox{$\,${\rm cm$^{\rm -#1}$}}\fi}
\def\raw {\ifmmode\rightarrow\else$\rightarrow$\fi}
\def\ex#1{\ifmmode {\times 10^{#1}}         
        \else \hbox{{$\times 10^{\rm #1}$}}\fi}

\newcommand{\kms}{\mbox{km~s$^{-1}$}}

\newcommand{\ls}{\mbox{$L_{\odot}$}}
\newcommand{\msun}{\mbox{$M_{\odot}$}}
\newcommand{\rsun}{\mbox{$R_{\odot}$}}

\newcommand{\ai}{\mbox{\'{\i}}} 
\newcommand{\vexp}{\mbox{$V_{\rm exp}$}} 
\newcommand{\vsys}{\mbox{$V_{\rm sys}$}} 
\newcommand{\vlsr}{\mbox{$V_{\rm LSR}$}}

\newcommand{\teff}{\mbox{$T_{\rm eff}$}}

\newcommand{\hal}{\mbox{H$\alpha$}}

\slugcomment{To appear in ApJ}

\shorttitle{Optical spectroscopy of evolved stars}
\shortauthors{S\'anchez Contreras et al.}

\begin{document}


\title{Echelle long-slit optical spectroscopy \\
                of evolved stars}


\author{C. S\'anchez Contreras\altaffilmark{1}, 
R. Sahai\altaffilmark{2}, A. Gil de Paz\altaffilmark{3}, and R. Goodrich\altaffilmark{4}}

\altaffiltext{1}{Departamento de Astrof\ai sica Molecular e Infrarroja, Instituto de 
Estructura de la Materia, CSIC, Serrano 121, E-28006 Madrid, Spain}

\altaffiltext{2}{Jet Propulsion Laboratory, MS 183-900, California
Institute of Technology, Pasadena, CA 91109, USA}

\altaffiltext{3}{Departamento de Astrof\ai sica, Universidad Complutense de Madrid, 
Avda. de la Complutense S/N, E-28040, Madrid, Spain}

\altaffiltext{4}{W. M. Keck Observatory, 65-1120 Mamalahoa Highway,
Kamuela, HI 96743, USA}







\begin{abstract}
We present echelle long-slit optical spectra of a sample of objects
evolving off the Asymptotic Giant Branch (AGB), most of them in the
pre-planetary nebula (pPN) phase, obtained with the ESI and MIKE
spectrographs at the 10\,m Keck\,II and 6.5\,m Magellan-I telescopes,
respectively. The total wavelength range covered with ESI (MIKE) is
$\sim$3900 to 10900\AA\ ($\sim$3600 to 7200\AA). In this paper, we
focus our analysis mainly on the \hal\ profiles. Prominent \hal\
emission is detected in half of the objects, most of which show broad
H$\alpha$ wings (with total widths of up to $\sim$4000\,\kms). In the
majority of the \hal-emission sources, fast, post-AGB winds are
revealed by P-Cygni profiles.
In $\sim$37\% of the objects \hal\ is observed in absorption. In almost
all cases, the absorption profile is partially filled with emission,
leading to complex, structured profiles that are interpreted as an
indication of incipient post-AGB mass-loss. The rest of the objects
($\sim$13\%) are \hal\ non-detections. We investigate correlations
between the \hal\ profile and different stellar and envelope parameters. All
sources in which \hal\ is seen mainly in absorption have F-G type
central stars, whereas sources with intense \hal\ emission span a
larger range of spectral types from O to G, with a relative maximum
around B, and also including very late C types.  Shocks
may be an important excitation/ionization agent of the close stellar
surroundings for objects with late type central stars. Sources with
pure emission or P Cygni
\hal\ profiles have larger $J-K$ color excess than objects with \hal\
mainly in absorption, which suggests the presence of warm dust near
the star in the former. The two classes of profile sources also
segregate in the IRAS color-color diagram in a way that intense
\hal-emitters have dust grains with a larger range of
temperatures. Spectral classification of the central stars in our
sample is presented. For a subsample (13 objects), the stellar
luminosity has been derived from the analysis of the
\ion{O}{1}\,7771-5\AA\ infrared triplet. The location in 
the HR diagram of most of these targets, which represent $\sim$30\%
of the whole sample, is consistent with relatively high final (and,
presumably, initial) masses in the range $M_{\rm
f}$$\sim$0.6-0.9\msun\ ($M_{\rm i}$$\sim$3-8\msun).

\end{abstract}


\keywords{stars: AGB and post-AGB, stars: mass loss, circumstellar matter,
ISM: jets and outflows, planetary nebulae: general} 



\section{Introduction} \label{intro}

Intermediate mass stars ($\sim$1-8\msun) evolve from the Asymptotic
Giant Branch (AGB) to the Planetary Nebula (PN) phase through a
short-lived ($\sim$\,10$^3$\,yr) and fascinating evolutionary stage
designated as the post-AGB (pAGB) or pre-planetary nebula (pPN) phase.  At some
point in the late-AGB or early pAGB stage, a process (or
processes) becomes operative that accelerates and imposes severe
asymmetries upon the slow, spherical AGB winds: the spherical, slowly
expanding ($V_{\rm exp}$\,$\sim$\,15\,\kms) AGB circumstellar envelope
(CSE) becomes a PN with clear departures from sphericity and fast
($\ga$\,100\,\kms) outflows directed along one or more axis.  Although
there is no consensus yet for what causes this spectacular
metamorphosis, fast jet-like winds have been hypothesized to play an
important role \cite[see e.g.\,the review paper on PNs shaping
by][]{bal02}. These outflows carve out an imprint within the AGB CSE
producing and shaping the fast, bipolar lobes observed in most pPNs
and PNs \citep{sah98}. The mechanism that powers and collimates
pAGB jets is a fundamental issue on stellar evolution that remains
a mystery.


Optical spectroscopic observations of pPNs and PNs have allowed
considerable advances in our understanding of pAGB evolution and, in
particular, are very useful for probing the interaction between pAGB
winds and the CSE formed in the previous AGB phase.  Our current (very
limited) knowledge of pAGB evolution and, more particularly, of pAGB
winds is derived mainly in two ways. The first is {\sl indirect}, that
is, based on the effects of the pAGB winds on the AGB CSEs.  Many
pPNs/PNs show extended lobes with, often, bow-shaped features at their
tips that are visible through optical recombination and forbidden
emission lines.  Spectroscopic observations have been crucial for
understanding the origin of these regions, which are found to be
excited by the passage of fast ($\ga$100\,\kms) shocks
\citep[e.g.][]{san00,vaz00}. From these results we infer the existence
of fast, pAGB winds that interact hydrodynamically with the AGB CSE
leading to the formation of shocks and, ultimately, to the
acceleration and shaping of the nebular material.

{\sl Direct} detection of pAGB winds is limited to a very few
objects. In some cases, there is Balmer and forbidden line emission
arising in a set of compact, shock-excited regions located along the
nebular axis \citep[e.g.][]{buj98,rie06}. These `knots', which usually
move away from the star at high velocity, are thought to result from
the propagation of shocks in the pAGB wind itself, suggesting that
pAGB winds are collimated and directed along the nebular axis. In some
pPNs/PNs, fast, pAGB winds are also revealed by P-Cygni profiles close
to the central star. One of the most exciting results recently
provided by long-slit spectra with $HST$/STIS is the first direct
observation of the spatio-kinematic structure of a fast collimated
pAGB wind (jet) in a `pristine' stage, i.e$.$, not strongly altered by
the interaction with the AGB shell \citep{san01}. The characterization
of the pristine jet in Hen\,3-1475,
which is collimated and expands at more than 2000\kms\,at only
$\approx$10$^{16}$\,cm from the central star, has had a direct impact
on the different theories of pAGB wind collimation, ruling out
purely hydrodynamical processes as the collimation agent in this
object.

Recognizing the importance of optical spectroscopy of pAGB objects we
have recently obtained optical long-slit spectra a large sample of
objects evolving off the AGB, most of them pPNs and young PNs, in
order to characterize the pAGB mass-loss process and the jet-sculpting
of AGB winds. This forms part of our extensive, multi-wavelength
survey program of imaging and spectroscopic observations of OH/IR
stars (evolved mass-losing stars with OH maser emission), which may be
the youngest pPNs \citep[e.g.][]{sahs07,san04}. 

In this paper, we present the spectroscopic database resulting from
our observations.  Our work complements recent optical spectroscopic
surveys of pPNs/PNs \cite[e.g.][]{sua06,per07} by providing deep, high
spectral resolution spectra covering a wide wavelength range.  Here,
we focus on the characterization and analysis of the different types
of H$\alpha$ profiles and their correlation with several stellar and
envelope parameters. 
The shape of the line profile is interpreted in terms of present-day
mass-loss in these objects, which enable obtaining information on
their on-going pAGB winds. We have also derived spectral types based
on the absorption line spectrum of our targets. Whenever available,
the \ion{O}{1}\,7771-5\AA\ infrared triplet have been used as a
luminosity indicator.




\section{Observations and data reduction}
\label{obs}

\subsection{Spectroscopy with ESI at the 10\,m Keck\,II telescope} \label{obsesi}

Observations of most targets (28 sources) were carried out in 2002 and
2003, using the Echellete Spectrograph and Imager (ESI; Sheinis et
al., 2002) in echelle mode mounted on the 10\,m W.M. Keck\,II
telescope at Mauna Kea (Hawaii, USA). Table\,\ref{t_log} presents a
journal of the observations.  The detector was a MIT-LL CCD with
2048$\times$4096 squared pixels of 15$\mu$m. Total wavelength coverage
is $\sim$3900-10900\AA. The reciprocal dispersion and the pixel
angular scale range from 0.16 to 0.30 \AA/pixel and from 0\farcs120 to
0\farcs168, respectively, for the ten echellette orders (15 to 6) of
ESI. The velocity dispersion has a nearly constant value of
11.5\kms\,pixel$^{-1}$ in all orders.  We used a
0\farcs5$\times$20\arcsec\ slit. For extended objects with known
morphology the slit was centered on the nucleus of the nebula and
oriented along its main symmetry axis. For the rest of the objects the
slit was oriented along the paralactic angle. For the object
IRAS\,05506$+$2414, we observed two different positions centered on
the compact sources labeled as $Sa$ and $Sb$ in the $HST$ images
presented in
\cite{sahs07}. Seeing ranged between $\sim$0\farcs6 and $\sim$0\farcs9. Weather
conditions were photometric only for the nights 2003, June 4 and 2004,
August 11.

All spectra were reduced using the IRAF\footnote{IRAF is distributed
by the National Optical Astronomy Observatories, which are operated by
the Association of Universities for Research in Astronomy, Inc., under
cooperative agreement with the National Science Foundation.} package
following the standard procedure.  The long-slit spectra were
bias-subtracted (using the IRAF task {\tt esibias}) and flat-fielded.
Individual exposures for each object were combined and
cosmic-rays removed. One dimensional (1D) spectra were extracted using
the IRAF task {\tt apall}. We used extraction apertures large enough
to include all the nebular emission (except for Hen\,3-1475; see
\S\,\ref{res}).  Background regions were defined on each side of the
aperture separated by a buffer zone for the different orders. The
background level in these regions was fitted and
subtracted. Wavelength calibration was done using exposures of
copper-argon (CuAr) calibration lamps and the night sky lines. The
comparison spectra were extracted using the same aperture as the
objects. Lines were identified in the calibration spectra and a
dispersion solution was found for, and applied to, each of our science
targets.
The velocity resolution achieved (FWHM of the lamp lines) is
$\sim$37\kms\ for all orders. Flux calibration was obtained by
observing a number of spectrophotometric standard stars (Hiltner 600,
Feige 56, LTT 1788, BD+28\degr4211, BD+33\degr2642, and Feige 110) using a
6\arcsec-wide slit.



\subsection{Spectroscopy with MIKE at the 6.5\,m Magellan-I telescope} \label{obsmike}

Spectra for IRAS\,17150$-$3224 and IRAS\,17440$-$3310 were obtained at
the 6.5-m Magellan-I (Baade) telescope in Las Campanas Observatory
(Chile) using the Magellan Inamori Kyocera Echelle (MIKE) spectrograph
\citep{ber03}.  (We also observed IRAS\,17441$-$2411 with MIKE,
however, the ESI spectrum for this source has a larger S/N and,
therefore, MIKE data are not presented.)  Observations were carried
out on 2003, April 27-28.  For the configuration used in
these observations, the wavelength ranges covered were
$\sim$3200--5000\,\AA\ and 4900--7300\,\AA, respectively for the blue
and red arms. Each arm is equipped with a 2048$\times$4096
15\,$\mu$m-pixels CCD. We used a 1\arcsec-wide slit, leading to a
resolving power of R$\simeq$28,000 (22,000) for the blue-arm (red-arm)
spectra.  The images were binned 2$\times$2 for a final reciprocal
dispersion of $\sim$0.04 and 0.10\,\AA\,pixel$^{-1}$, respectively for
the blue and red arms, and a spatial scale of
$\sim$0.28\,arcsec\,pixel$^{-1}$.  Since our targets were more
extended that the length of the slit projected onto the sky
(5\arcsec), blank sky spectra with the same exposure time as for our
source were taken right after observing each object for background
subtraction. The average seeing was $\sim$1\arcsec\ ($\sim$1\farcs5)
for the night when IRAS\,17150$-$3224 (IRAS\,17440$-$3310) was
observed.

For IRAS\,17440$-$3310 and IRAS\,17150$-$3224 the slit was centered on
the nebular nucleus and the North-West lobe, respectively. The PAs
along which the slits were oriented rotated during the exposure
$\pm$5-7\degr\ around the position quoted in Table\,\ref{t_log}.  This
is because MIKE is located in one of the Nasmyth platforms of the
Magellan telescope but is not attached to the instrument rotator
neither it incorporates any mechanism to compensate for the field
rotation. This setup provides an excellent stability but implies that
the position of the slit in the sky changes continuously as the
observation progresses.


These spectra were reduced using the MIKE pipeline developed by Daniel
Kelson (see Kelson et al$.$ 2000, Kelson 2003, 2007 for an extensive
description of some of the numerical methods used). Although
observations were performed under non photometric conditions, we
corrected the data for the wavelength-dependent response of the system
by acquiring spectra of the spectrophotometric standard stars HR 3454,
HR 4468, HR 7596, HR 8634, and HR 7950. All the spectra were
wavelength calibrated using observations of a lamp of Thorium-Argon
taken right after observing each target. The velocity resolution
achieved (FWHM of the lamp lines) is nearly constant for all the
orders: $\sim$12-13\kms\ for the red arm and $\sim$10\kms\ for the
blue arm.  One dimensional spectra were extracted using the total
length of the slit as the aperture. Our spectra probably do not
include emission from the ends of the lobes of IRAS\,17150$-$3224 and
IRAS\,17440$-$3310, which lied outside of the slit.







%







\subsection{The sample} 
\label{sample}

The objects observed in this work are listed in Tables \ref{t_log} and
\ref{t_parms}.  In Table \ref{t_parms} we give the type of \hal\,
profile assigned in this study (\S\,\ref{res_hi}) together with
the class of object, the spectral type of the central star, the
f12/f25 IRAS flux ratio, the 60\micron\,IRAS flux (f60), the
morphology of the optical and/or near-IR nebula, and the chemistry.
We have adopted the primary morphology classification system by
\cite{sahs07}, which establishes four main classes of nebular shapes:
bipolar (B), Multipolar (M), elongated (E), and irregular (I).  We
have also included information on two secondary structural features
defined by Sahai et al., namely, the presence of a dark obscuring
waist across the center of the nebula (denoted by ``w'') and direct
visibility of the central star in the optical or near-infrared
(denoted by ``$\star$''). Objects with star-like appearance in the
$HST$ images, i.e$.$\ with no nebulosity detected around them, are
denoted as stellar (S).  References for the spectral type, published
high-angular resolution images (optical and/or near-IR), and chemistry
are given in the last column of Table\,\ref{t_parms}.  Many of the
objects in our sample are listed in the ``The Torun catalogue of
Galactic post-AGB and related objects'' by \cite{szc07}, where
additional information and references can be found.

Our sample has been mainly selected from our multi-wavelength surveys
of OH/IR stars \citep[e.g.][]{sahs07,san04} based on their IRAS colors and identification of
optical counterparts.  OH/IR stars generally show double-peaked
OH-maser emission lines profiles, are strong infrared sources, and are
believed to be evolved mass-losing stars with dense CSEs. The IRAS
spectral energy distributions (SEDs) of most of our targets show a lack of
hot dust ($F_{12}$\,$<$\,$F_{25}$) -- indicating a recent cessation of
the large-scale AGB mass-loss process, which is believed to signal the beginning
of pAGB evolution.
The complex, aspherical morphologies of most objects as seen in
high-angular resolution images indicate that pAGB wind
interactions are taking, or have taken, place in these sources.

Although our sample is composed in its majority of pPNs, it also
includes five long-period variable (AGB) stars: IRAS 01037+1219 (IRC+10011), IRAS
02316+6455 (V 656 Cas), IRAS 03507+1115 (IK Tau), IRAS 09452+1330 (IRC+10216) and IRAS
10131+3049 (CIT 6). Their IRAS colors ($F_{12}$\,$>$\,$F_{25}$)
indicate that the heavy AGB mass-loss process is still going
on. The large-scale envelopes around these objects are
roughly circularly symmetric, however, the core nebular structures are
often asymmetric and display bipolar or irregular/clumpy
morphologies. This indicates that these objects are likely precursors
of bipolar pPNs in which the jet-sculpting of the AGB CSE by interaction with 
collimated (probably fast) outflows has already started.
Our sample also contains several young PNs (yPNs).  We have
adopted the classification as yPNs for objects with O- and B-type
central stars surrounded by newly formed, compact \ion{H}{2} regions
like the well known pAGB objects M\,1-92 and Hen\,3$-$1475 (see
Table\,\ref{t_parms}). We have also included in our sample the object
IRAS\,19114$+$0002 (also known as AFGL\,2343), which has a
controversial evolutionary status.  IRAS\,19114$+$0002 has been
classified as either a pPN descended from low- or intermediate-mass
progenitor \citep{rh99} or a yellow hypergiant (YHG) descended from a
massive ($\sim$30\msun) Population I supergiant
\citep[e.g.][]{jur99,cas07}.  YHGs are believed to follow an
evolutionary path similar to that of pPNs: they undergo a strong
mass-loss process leading to massive CSEs while their central stars
become progressively hotter, therefore, we consider appropriate to
include IRAS\,19114+0002 in our observations.  The source
IRAS\,05506$+$2414, although satisfies our selection criteria, is most
likely not an evolved star. This enigmatic outflow source was
serendipitously discovered as part of our multi-wavelength survey of
pPNs and is probably associated with an early stage of
a massive star's life \citep{sah08}. The ESI spectra of
IRAS\,05506$+$2414 taken at position $Sa$ (see \S\,\ref{obsesi}) is
presented here for completeness but this object is not further
discussed except for spectral type assignation
(\S\,\ref{specclass}). Finally, although most of the objects in this
study are O-rich, we have also incorporated some C-rich envelopes as well
as objects with a mixed or uncertain chemistry.

Our sample is not complete on any of the properties summarized in
Table \ref{t_parms}, but rather should be considered as a
pilot sample. As happens in other studies based on optical data, our sample
is necessarily biased toward objects with optical counterparts. This
implies that we may be missing pAGB objects heavily obscured by thick,
dusty envelopes, i.e.\ probably the most massive and youngest pAGB
stars. What is the fraction of such massive, heavily obscured objects
and what is the mass/age threshold in our sample is very difficult to
determine because the star and envelope evolution (as
well as their mutual influence) along the AGB and pAGB phases, which
ultimately determine when the envelope becames visible in the optical,
remain very poorly characterized.

\section{Results} 
\label{res}

For most of the targets, our two-dimensional (2D), long-slit spectra
show spatially unresolved or marginally resolved emission. 
For the sources IRAS\,05506$+$2414, IRAS\,17423$-$1755,
IRAS\,19343$+$2926, IRAS\,21282$+$5050, IRAS\,22036$+$5306,
IRAS\,17317$-$2743, IRAS\,17440$-$3310, IRAS\,17441$-$2411, and
IRAS\,17150$-$3224, the nebular emission is resolved. The 2D spectra
of these spatially resolved nebulae will be studied in detail in the
future. In this paper, we present and discuss extracted 1D spectra
obtained using an aperture that includes emission from the whole
nebula. In most cases, the spectrum is dominated by that of the bright
nucleus, the emission from the extended nebulosity being undetected or
very faint. In the case of Hen\,3-1475, the aperture has been chosen
to include emission only from the bright nebular core (inner 2\arcsec)
and leaving outside the emission from the shocked, axial knots that
are well separated from the nucleus \cite[e.g.][]{san01}.  For the
objects in our sample with optically thick equatorial regions (a total
of 8; see references given in Table\,\ref{t_parms}), the light from
the nucleus, including the stellar continuum, is indirectly seen
scattered off by the dust in the lobes.

The spectra of our targets are composed by a number of absorption
and/or emission lines superimposed on a fairly red continuum
(Fig.\,\ref{f_fullspec}). Most absorption lines are expected to be of
photospheric origin whereas emission (recombination and forbidden)
lines are most likely nebular. Other absorption features that may have
an interstellar or circumstellar origin, such as the diffuse bands at 4430, 5780, 5797,
and 6284\AA\ \cite[e.g.][and references therein]{luna08}, are also
observed in some of our targets.
The continuum emission is dominated by the stellar photospheric
continuum (we have checked that the contribution by nebular continuum
is negligible, as expected, even for the objects displaying the most
intense \hal\ emission).

\subsection{Hydrogen lines}
\label{res_hi}

Half of the objects in our sample (15 out of 30) show \hal\ mainly in
emission (Fig.\,\ref{fig1-em}) and 11 targets show \hal\ mainly in
absorption (Fig.\,\ref{fig1-efA}). We have also found 3 sources (all
of them are AGB stars) with neither
\hal\ emission or absorption  and 1 object (the pPN IRAS\,19292$+$1806) 
in which the low S/N of the spectrum prevents us from determining
whether \hal\ emission or absorption is present.

Based on the \hal\ profile we can identify objects of different types
(Table\,\ref{t_parms}). Among the sources
where \hal\ is observed mainly in emission (Fig.\,\ref{fig1-em}), we find two basic types of
profiles, namely, symmetric and asymmetric. Only 2 sources belong to
the first category (referred to as pure emission sources -- {\bf pE}):
the yPN IRAS\,19374$+$2359 and the AGB star IRAS\,09452$+$1330
(IRC$+$10216). In CIT 6 there is strong line blending around the
\hal\ line\footnote{We note that the features around \hal\ are not due to the noise but represent 
real emission lines, many of which are also observable in the spectrum
of IRC$+$10216.}; since \hal\ cannot be isolated, determining
the shape of its profile is not possible. The rest of the
\hal-emitting sources, which is also the majority (12 out of 15), show
signs (in a different manner and amount) of absorption bluewards of
the line core, which is responsible for their asymmetric profiles.
Most sources show clear P Cygni like profiles, i.e$.$\ with a
narrow/sharp blue-shifted absorption feature against the line
wings. IRAS\,08005$-$2356 and IRAS\,19024$+$0044 display two of the
most remarkable P Cygni profiles amongst the objects in this category
(referred to as {\bf pcyg}). In IRAS\,19520$+$2759 (and tentatively
IRAS\,17516$-$2525 and IRAS\,19306$+$1407), not one but two different
blue-shifted absorption features are observed. In 3 objects (He3-1475,
IRAS\,21282$+$5050, and IRAS\,20462$+$3416), the
\hal\ blue wing is clearly weaker than the red one. Although a sharp
blue-shifted absorption is not observed in our spectra, these objects
belong to the pcyg class. In the case of the well known yPN
Hen\,3$-$1475, $HST$/STIS optical spectra show a remarkable \hal\
P Cygni profile with two distinct blue-shifted absorption features
well delineated against the line wings (S\'anchez Contreras \& Sahai
2001). These absorption features appear smoothed out in our ESI
spectra and, more generally, in ground-based observations probably due to their
blending with emission components arising in different nebular regions
superimposed within the PSF. In the case of IRAS\,21282$+$5050 and
IRAS\,20462$+$3416, P Cygni like profiles are observed in other
\ion{H}{1} and \ion{He}{1} lines (see e.g.\ H$\beta$ in
Fig.\,\ref{f-hb}), which supports \hal\ having a true, but partially
masked out, P Cygni profile. Moreover, further
confirmation of P Cygni line profiles exists from previous
spectroscopic observations of both IRAS\,21282$+$5050 
and IRAS\,20462$+$3416 \cite[][and references therein]{smi94,arr03}.


In most of the objects showing \hal\ absorption
(Fig.\,\ref{fig1-efA}), the line profile is found to be partially
filled with emission (see a detailed fitting of the
different absorption and emission components of the profile in
\S\,\ref{proffit}.) We refer to this objects as emission filled
absorption sources ({\bf efA}). The emission component is quite
prominent, reaching or exceeding the continuum level, for example, in
IRAS\,17440$-$3310, IRAS\,19114$+$0002, and IRAS\,19475$+$3119.  In
other cases, the \hal\ absorption shows a peculiar, structured profile
with an emerging emission feature or ``hump'' bluewards of the narrow,
deep absorption core: e.g.\, IRAS\,04296$+$3429, IRAS\,17441$-$2411,
and IRAS\,18167$-$1209.  Such complex emission-absorption \hal\
profiles are most likely the result of the superposition of two lines
of different origin: a photospheric absorption core and an emission
component that originates in the stellar close surroundings (see
further discussion in \S\,\ref{dis_efA}).  There are only two objects
in our sample, IRAS\,17150$-$3224 and IRAS\,19477$+$2401, for which
the \hal\,line presents a pure absorption ({\bf pA}) profile, at
least, within the limited S/N of our data.  In a number of efA and pA
objects, the \hal\ absorption profile has two components: a deep,
narrow central absorption feature and broad absorption wings. Similar
\hal\ profiles are typically found in F-type supergiants, in which the
broad wings are most likely due to the Stark broadening mechanism
operating at deep layers of the stellar atmosphere, whereas the narrow
component is consistent with being formed in the low surface gravity
photosphere of these stars.
The blue emission-hump observed in the absorption profile of
some objects is also commonly found in F-type supergiants.

We briefly discuss now the profile of other recombination \ion{H}{1}
lines, whenever they are detected, and compare it with that of
\hal. As expected, none of the 3 AGB stars with \hal\ non-detections 
show other \ion{H}{1} lines either in absorption or emission. In the
pPN IRAS\,19292$+$1806, which is not detected in \hal\ because of low
S/N in that spectral region, weak Paschen lines (from Pa$_{16\rightarrow3}$ at
8502\AA\ to Pa$_{9\rightarrow3}$ at 9229\AA) are seen in absorption. For pA and
efA objects whenever other lines of the Balmer or Paschen series are
detected they show pure absorption profiles --- i.e$.$\ the incipient
emission seen in \hal\ for efA objects is not visible in other
\ion{H}{1} lines.
Among the two objects for which \hal\ is
observed with a pure emission profile there are no other \ion{H}{1}
lines detected except for CIT 6, which shows H$\beta$ emission with a
quite broad, most likely blended profile. 
Except for IRAS\,17516$-$2525 and IRAS\,22574$+$6609 (see below), all
sources with \hal\ P Cygni profiles display the same type of profile
in one or more of the Balmer lines and also, in some cases, in the
Paschen series (observable in our ESI spectra from the Paschen
discontinuity at 8200\AA\ to Pa$_{7\rightarrow3}$ at 10049.4\AA; note,
however, that the Pa lines from upper levels $n$=16, 15, and 13 at
8502, 8545, and 8665\AA, respectively, are usually blended with the
near-IR \ion{Ca}{2} triplet and determining their profiles
unambiguously is difficult).
The P Cygni profile is more easily recognized and remarkable in
H$\beta$ than in \hal, in which the intense emission wings partially
infill the absorption component (see e.g.\ IRAS\,21282$+$5050;
Figs.\ref{fig1-em} and \ref{f-hb}).  The absorption component of the P
Cygni profile appears more prominent (sharper and deeper) relative to
the emission for bluer Balmer lines.
This is partially an effect of the emission component becoming
progressively weaker for the recombination lines from higher levels
(which are expected to be less and less populated as the quantum
number $n$ increases). In some cases, the emission component almost
disappears completely for the bluest Balmer lines observed by us
(H$\gamma$ and H$\delta$), resulting in a blue-shifted pure
absorption profile. 
In contrast to the emission component, the absorption strongly depends
on the population of level $n$=2, which is the same for all the Balmer
lines. IRAS\,17516$-$2525 and IRAS\,22574$+$6609 are the two only pcyg
sources for which Pa lines are observed with a pE and pA profile,
respectively, i.e$.$\ non P Cygni. (In both cases, the S/N in the
spectral region of the Balmer series is too low to determine whether or
not these lines are present.) 

\subsubsection{Variability of the \hal\ profile} 
\label{var_hal}

We have noticed differences in the \hal\ profiles of some of our
targets with respect to earlier observations. Here, we describe and
briefly discuss such differences for compact (unresolved or marginally
resolved) objects in which the observed variations most likely
represent real changes, e.g$.$ are unlikely to be due to the different
slit width and orientation in our spectra and earlier datasets.


{\it IRAS\,19114$+$0002}. Our \hal\ profile shows a narrow absorption
feature ($\lambda_{\rm obs}$=6564.5\AA) and two adjacent (blue- and
red-shifted) emission components (Fig.\,\ref{fig1-efA}).  There is
another absorption feature bluewards of \hal\ ($\lambda_{\rm
obs}$=6561.1\AA) that is most likely the \ion{Ti}{1}
$\lambda$6559.57\AA\ line as suggested by \cite{zac96}. In fact, the
center of the \hal\ absorption component and that attributed to the
\ion{Ti}{1} line yield a very similar value of the systemic velocity,
\vlsr=111 and 104\,\kms, respectively, in good agreement with earlier
results from other absorption lines and CO measurements
\citep[e.g.][and references therein]{rh99,buj01}.
\cite{zac96} also observed a double-peaked emission component and a 
narrow absorption core in the \hal\,profile of this object, i.e$.$\ a
shell profile, however, in our spectrum the red emission component is
stronger than the blue one, in contrast to what these authors
found. Moreover, the velocity difference between the two emission
peaks measured by us, $\sim$145\,\kms, is larger than that observed by
\cite{zac96}, $\sim$128\,\kms.

{\it IRAS\,19475$+$3119}. Variations of the \hal\ profile observed
with ESI with respect to previous measurements are reported by 
\cite{sah07}. Our \hal\ profile shows a broad absorption feature and a
narrow inverse P Cygni shaped core (Fig.\,\ref{fig1-efA}).  The \hal\
profile observed by \cite{klo02} shows similar broad absorption wings,
however, the line core shows two emission peaks neighboring 
a deep, narrow absorption feature. \cite{klo02} also present and compare
their \hal\ profiles for three different epochs and show that the
intensity of the red (blue) emission peak gradually increases
(decreases) with time.
The line-shape observed by us fits then quite well with the observed
trend, representing the extreme case in which the blue peak has
disappeared completely.  

{\it IRAS\,20462$+$3416}. Our \hal\ profile is different from that
previously reported by \cite{smi94} and \cite{gl97}, in particular,
the blue-shifted absorption component of the P Cygni profile is less
pronounced in our observations.
These authors already noticed strong changes in the profile of
\hal\ and other lines in a time-scale of days, which are interpreted in 
terms of mass-loss episodes \citep[see also][]{ark01}.  We confirm
such changes in some recombination lines, for example, we detect the
\ion{He}{1}$\lambda$6678\AA\ line in absorption, however, this line
has had a remarkable P Cygni profile in the past.


{\it AGB stars: IRC+10216, CIT\,6, and IK\,Tau.}  Line emission in
long-period variable stars is known to be a transient phenomenon.
\hal\ emission was not present in the spectrum of the C-rich
variable star IRC$+$10216 observed by \cite{tram94} but was detected by
\cite{coh82} as well as in this work. The other C-rich AGB star in our
sample, CIT\,6, has gone through emission-line phases showing intense
Balmer line emission and other non-hydrogen lines like [\ion{O}{1}],
[\ion{N}{2}], and [\ion{S}{2}] suggestive of emission from a
low-excitation shocked zone \citep{coh80,coh82,tram94}.  In our ESI
spectrum, \hal\ appears much weaker than in earlier works and is
blended with other (presumably metallic) lines with similar intensity.
Finally, we do not detect \hal\ or any other emission lines in the
O-rich Mira variable IK\,Tau, however, this object must have shown
line emission in the past according to its M6e-M10e assignation in the
General Catalogue of Variable Stars (GCVS). 


\subsection{Non-Hydrogen line emission spectrum}
\label{res_nohi}

\begin{figure}
\plotone{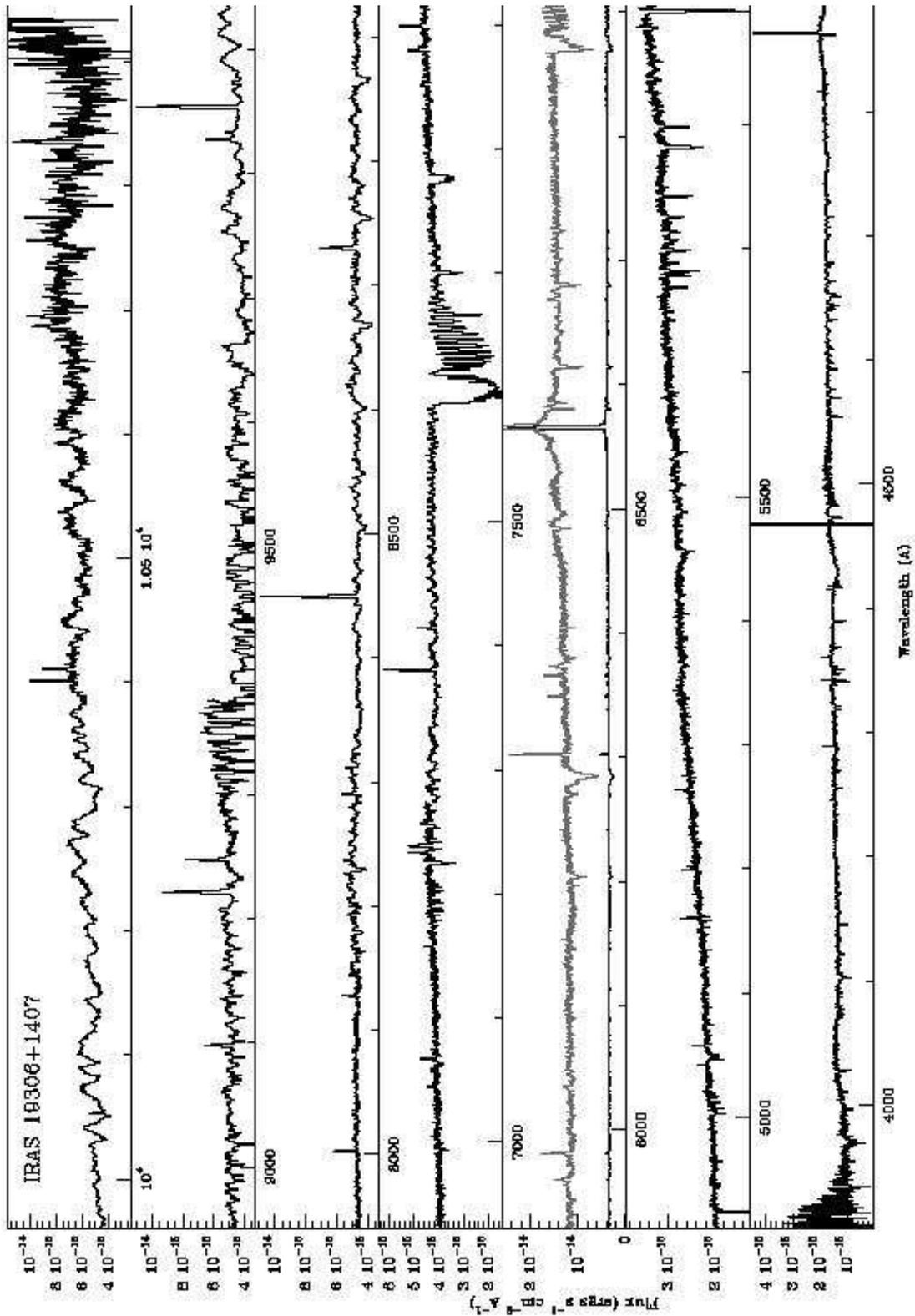}
\caption{The spectrum of IRAS\,19306$+$1407 observed with ESI is shown here as an example. In some panels, the spectrum is also plotted using a smaller flux scale for clarity (grey line).
Spectra for all sources are available in the electronic edition of the journal. \label{f_fullspec}} 
\end{figure}

\begin{figure}
\epsscale{0.8}
\plotone{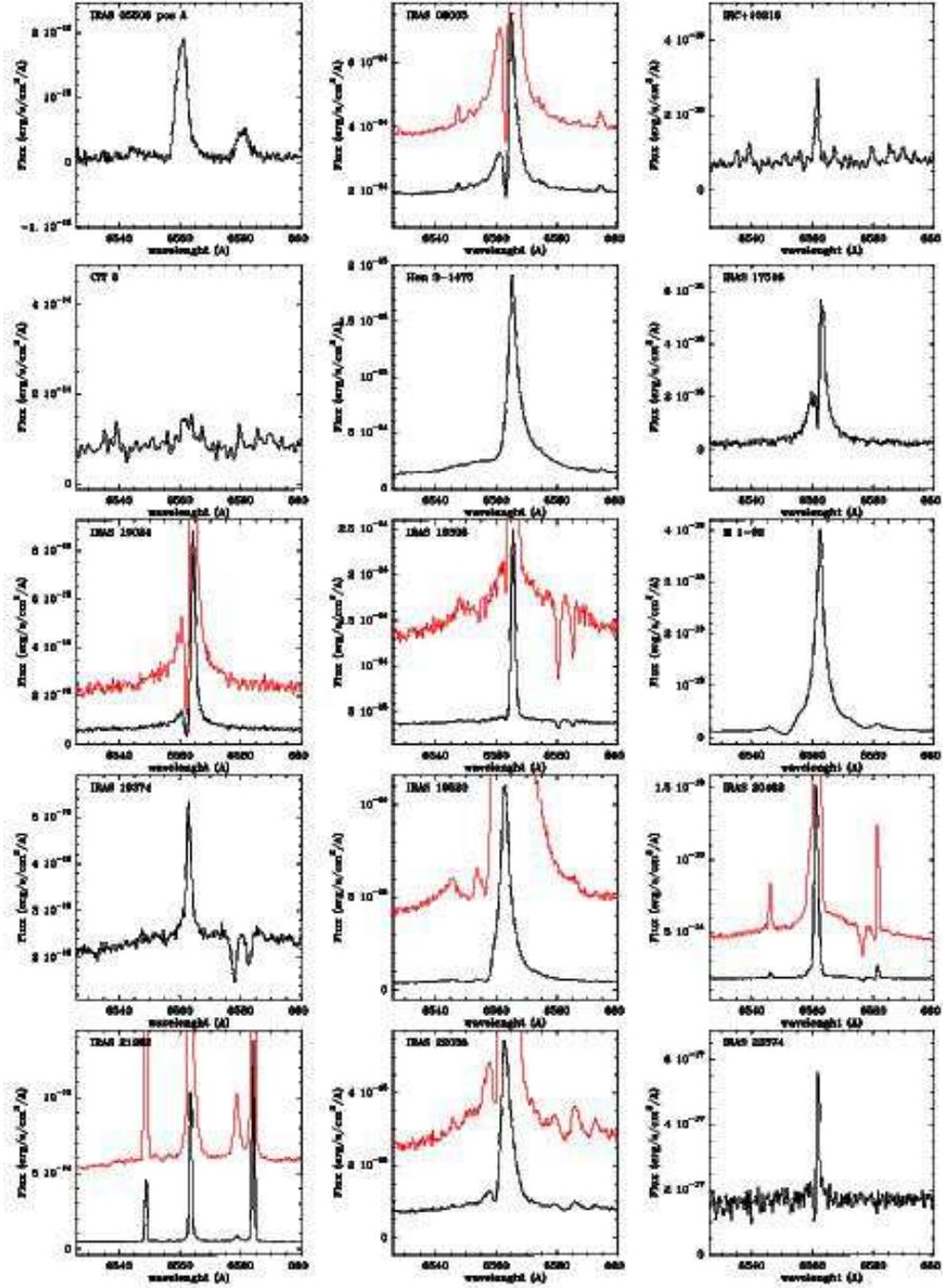}
\caption{Spectra in the region around the H$\alpha$ line for objects
  with H$\alpha$ emission, i.e$.$\ pcyg and pE sources
  (Table\,\ref{t_parms}). In some cases, the spectrum is also plotted
  using a smaller scale for clarity (red line).
\label{fig1-em}}
\end{figure}

\begin{figure}
\plotone{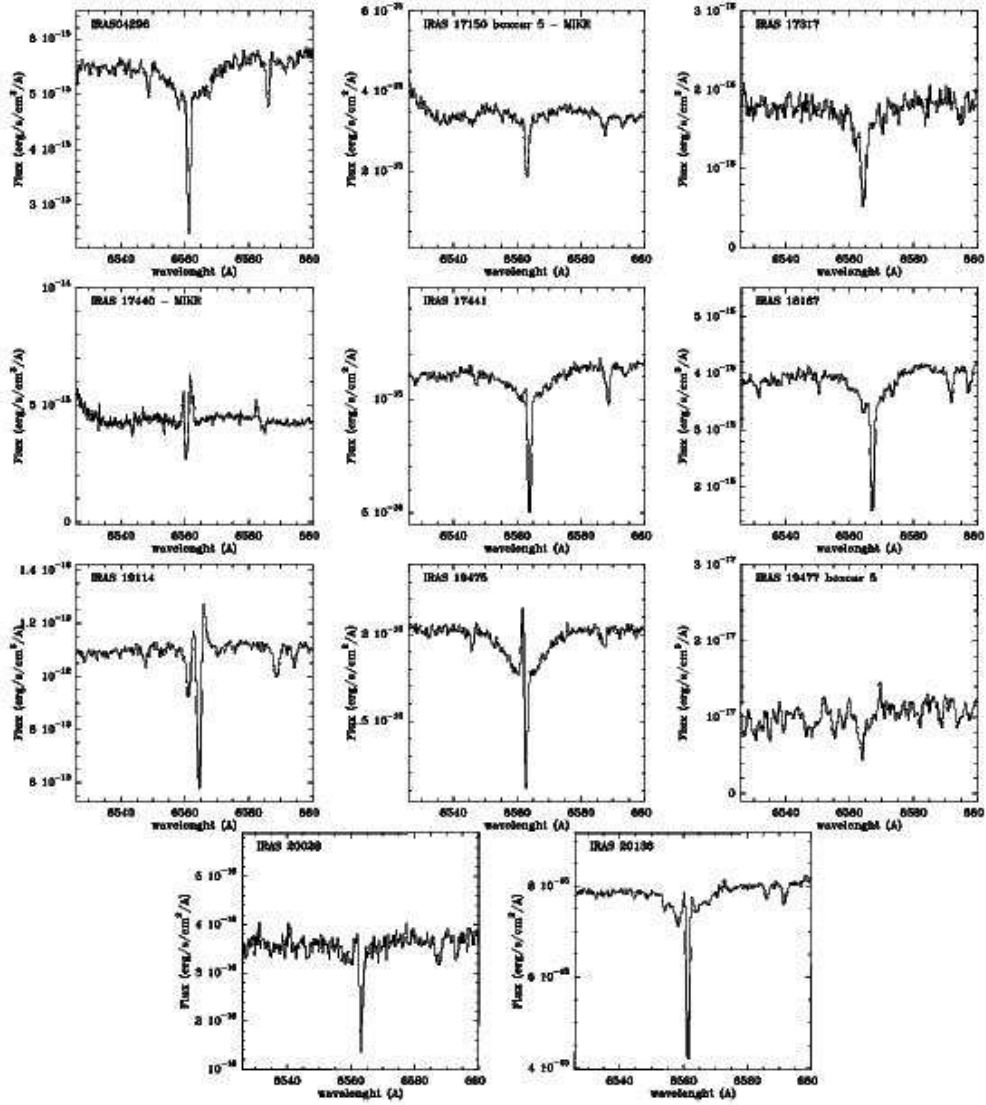}
\caption{Spectra in the region around the H$\alpha$ line for objects
  with H$\alpha$ absorption, i.e$.$\ efA and pA sources
  (Table\,\ref{t_parms}). The spectra of IRAS\,17150$-$3224 and 
  IRAS\,19477+2401 have been
  boxcar smoothed (using a flat-topped kernel of 5 pixels) to increase
  S/N. \label{fig1-efA}}
\end{figure}

\begin{figure}
\plotone{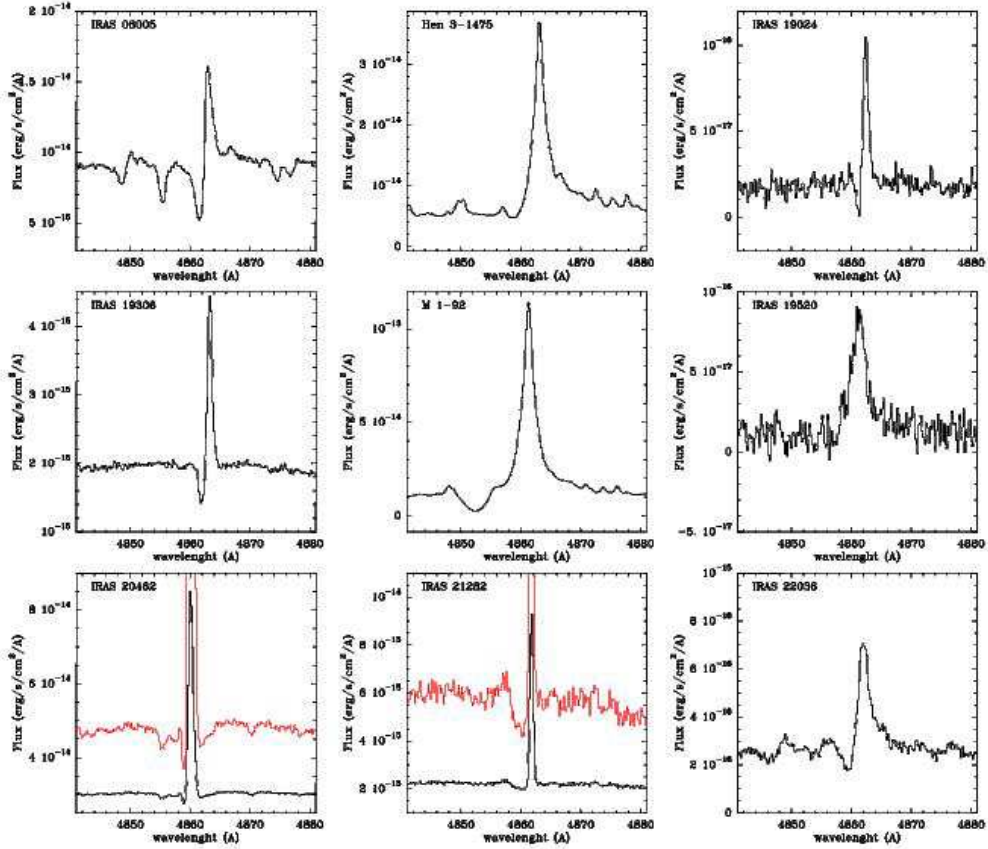}
\caption{Spectra in the region around H$\beta$ for sources with \hal\
  P Cygni like profiles (Table\,\ref{t_parms}).  The spectra of the
  pcyg sources IRAS\,17516$-$2525 and IRAS\,22574$+$6609, for which
  signal-to-noise ratio is $<$3 in this region are not plotted.  For
  IRAS\,20462$+$3416 and IRAS\,21282$+$5050, the spectrum is also
  plotted using a smaller scale for clarity. \label{f-hb}}
\end{figure}

Most objects in our sample with intense \hal\ emission (i.e$.$ pcyg
and pE) also exhibit other emission lines by heavier elements. In two
cases, namely, the yPNs M\,1-92 and Hen\,3-1475, the underlying
stellar absorption spectrum is totally masked out by a wealth
(hundreds to thousands) of emission lines by different neutral and
ionized atoms, like \ion{He}{1},
\ion{N}{1}, \ion{N}{2}, \ion{C}{1}, \ion{O}{1}, \ion{O}{2},
\ion{S}{2}, \ion{Fe}{1}, \ion{Fe}{2},... (see full spectra of our
targets in Fig.\,\ref{f_fullspec}) --- a comprehensive list of nebular
emission lines identified in the optical spectrum of M\,1-92 (many of
them also detected in Hen\,3-1475) is reported by \cite{arr05}.
The rest of the \hal-emitting sources in our sample, which is also the
majority, show a small or moderate number of emission lines, lacking
many of the intense forbidden emission lines typical of evolved pNe
with hot (\teff$\gg$30,000K) central stars
\cite[e.g.][]{vds96a,vds96b}. 
Although the spectrum of the two AGB stars in our sample with \hal\
emission, IRC+10216 and CIT\,6, is clearly shaped by molecular
(e.g.\,VO, TiO, and ZrO) absorption bands and metallic absorption
lines typical of C-rich late type stars, several weak emission lines
are also detected. Hydrogen and metallic emission lines are known to
be present in the spectrum of long-period variables at certain phases
of their pulsational cycle \cite[e.g.][and references
therein]{cas00}. Such emission lines are interpreted as due to heating
of the stellar atmosphere by shock waves.  In these two objects we
also detect very prominent C$_2$ Swan bands $\lambda$6059 (2,4),
$\lambda$6122 (1,3), and $\lambda$6191 (0,2). To our knowledge, this
is the first time that these particular bands are reported in these
objects. 
In the spectrum of CIT\,6, we also detect the band $\lambda$5097 (2,2)
and tentatively $\lambda$5070 (3,3), $\lambda$5635 (0,1), and
$\lambda$5585 (1,2).  The Swan bands of the C$_2$ molecule are
commonly observed in C-rich stars and comets \citep[see e.g.][and
references therein]{klo99,bie06}.




Most sources in which \hal\ is observed in absorption, i.e$.$\,with a pA
or efA profiles, have a spectrum dominated by absorption lines with no
hint of nebular emission lines. The central stars of pA and efA
objects all have F and G spectral types, therefore, the lack of
emission lines is consistent with these relatively cool stars not
being able to ionize a significant fraction of their circumstellar
material. 
The are only three objects in which \hal\ is seen in
absorption that also exhibit weak emission features:
IRAS\,04296$+$3429, IRAS\,19114$+$0002, and IRAS\,20136$+$1309. (We
also find tentatively one weak emission line at $\sim$6604\AA\ in the
spectrum of IRAS\,19475$+$3119, which could be identified with the
very low excitation line \ion{Ni}{1}$\lambda$6604.29\AA.)



Two relatively intense emission features around 5632 and 5583\AA\ are
observed in the compact (unresolved) nuclear region of
IRAS\,04296$+$3429 (Fig.\,\ref{f_fullspec}). These lines, which are
very broad (FWZI$\sim$5\AA) and have a peculiar, triangular profile,
are identified as the emission bands (0,1) and (1,2) of the Swan
system of the C$_2$ molecule at 5635 and 5585\AA, respectively.  The
(0,0) $\lambda$5165, (1,1) $\lambda$5129, (0,2) $\lambda$6188
C$_2$ bands are also detected with a smaller S/N in our spectrum. The
(0,1) and (0,0) Swan bands have been previously reported by
\cite{klo99}, whereas the other three vibrational transitions are
reported by us for the first time. The emission band (1,0)
$\lambda$4735 is absent in our spectrum as well as in that presented
by \cite{klo99}. \cite{hriv95}, however, report detection of this band
in {\sl absorption} from low-resolution spectra. We confirm the
presence of a weak emission feature around 4071\AA\ previously noticed
by \cite{hriv95}. This line roughly coincides in wavelength with an
emission feature (attributed to a blend of [\ion{S}{2}] lines)
observed in the C-rich pPN CRL\,2688.  There are two other
unidentified emission lines in the spectrum of IRAS\,04296$+$3429 at
8065.6\AA\ and 8102.9\AA.
The narrow profile of these emission features is consistent with being
atomic or ionic lines rather than molecular bands.

There are only 2 emission lines (other than \hal) observed in the
spectrum of IRAS\,19114$+$0002 that we identify as the [\ion{Ca}{2}]
$\lambda\lambda$7291.47,7323.89\AA\ doublet. These lines, which arise
in the compact (unresolved) circumnuclear region, are also observed in
other objects in our sample: IRAS\,17516$-$2525, M\,1-92, Hen\,3-1475,
IRAS\,22036$+$5306, and IRAS\,08005$-$2356, and tentatively in
IRAS\,19520$+$2759. (We can't rule out or assert the presence of these
lines in the spectra of IRC+10216 and CIT 6, which display many
emission lines strongly blended in that region). Since calcium has
among the highest gas phase depletions, detection of these lines
indicate that calcium is not depleted onto dust grains either
because the [\ion{Ca}{2}] doublet emission arises in a dust-free
environment or because Ca is liberated from the grains. We consider
grain destruction by moderate-velocity shocks a likely explanation in
the case of IRAS\,19114$+$0002 and, more generally, for the pPNs and
yPNs in our sample with [\ion{Ca}{2}]-emission \cite[see
e.g.][]{hart87}. Moderate speed motions are in fact deduced from the
FWHM of the [\ion{Ca}{2}] doublet in our targets, which range between
40 and 100\kms. These values are larger than the typical expansion
velocities of AGB CSEs and probably result from shock-acceleration
produced by the pAGB-to-AGB wind interaction (\S\,\ref{intro}).
It is worth mentioning that the [\ion{Ca}{2}] doublet is also observed in the young stellar object
IRAS\,05506, which shows a low-excitation shock spectrum \citep{sah08}. 
The critical density for the [\ion{Ca}{2}] doublet is $\sim$10$^6$\cm3, 
therefore, emission of these lines arises in a relatively tenuous gas.

We detect a few weak emission lines in IRAS\,20136$+$1309
(Fig.\,\ref{f_fullspec}).  Some of these features are also present in the
spectra of the yPNs M\,1-92 and Hen\,3-1475 and the pPNs
IRAS\,08005$-$2356 and IRAS\,22036$+$5306, for example, the emission
around 5955 and 7911, which we tentatively identify with
\ion{Fe}{1}\,$\lambda$5956.7\AA\ and \ion{Fe}{1}\,$\lambda$7912.8\AA, respectively, or the lines
7083 (\ion{Ti}{1}\,$\lambda$7084.2\AA\ or
\ion{C}{1}\,$\lambda$7085.5\AA?) and 8045\AA\
([\ion{Cl}{4}]\,$\lambda$8045.6\AA\ or
\ion{Mg}{1}\,$\lambda$8047.7\AA?). 

\section{Analysis}
\label{anal}

\subsection{Parameter analysis of the \hal-profile}
\label{proffit}

The presence of \hal\ emission from the compact nebular core
displaying either a pE, pcyg, or a structured efA profile is
interpreted as an indication of on-going (i.e.\,pAGB) mass-loss most
likely in the form of a stellar wind (see also \S\,\ref{discuss}). In
this section, we parameterize and analyze the observed line profiles,
which is needed to derive information on the current stellar wind and
other processes that may be affecting the observed line shape.

\subsubsection{Pcyg and pE sources}
\label{prof_pcyg}

The parameters used to describe the H$\alpha$ profile for pcyg and pE
sources are given in Table \ref{t_pcyg}.  The
emission component of the \hal\ profile consists of an intense core
plus weak broad wings. 
In pcyg sources, the blue emission wing is affected by absorption,
therefore, in order to estimate the full width at half maximum (FWHM)
of the emission and absorption components separately in these cases,
we have fitted a Lorentz profile to the line emission core and wings.
The observed \hal\ profile has been subtracted from the Lorentzian
fit to retrieve the absorption line-shape. In doing so, we are
implicitly assuming that the {\sl intrinsic} emission profile as
produced in the nebula nucleus (i.e.\,before being altered by the
absorption) is symmetric.
Except for the FWHM of the absorption and emission lines, the rest of
the parameters in Table\,\ref{t_pcyg} have been measured directly on
the observed profile 
to avoid uncertainties resulting from the quality of the fit and/or
the validity of our assumption of an intrinsic, symmetric 
emission profile. Accordingly, the equivalent widths, $W_\lambda$, given in the
table\footnote{Although $W_\lambda$ is normally defined to be positive
(negative) for an absorption (emission) line, we express
both the absorption and emission $W_\lambda$ as a positive value for
simplicity.}  represent the total (emission plus absorption)
equivalent widths.
For pcyg sources, $W_\lambda$ is then a lower limit to the equivalent
width of the intrinsic nuclear emission.  Given the typical depth of
the absorption feature, the emission $W_\lambda$ is expected to be
underestimated only by a small factor (less than 2).  The parameters
of the absorption component $V$ and $V_{\rm max}$ are, respectively,
the Doppler shift of the centroid and the blue edge of the absorption
feature relative to the emission peak, which is located at LSR $V_{\rm
e}$, in
\kms. The values of the full width at
zero intensity level (FWZI) depend on the noise level in our spectrum
and, therefore, they must be regarded as lower limits.
Among the three objects with pE \hal\ profiles, the presence of broad
wings is only confirmed for IRAS\,19374$+$2359. In this case, however,
the total width of the wings is uncertain partially due to the limited
S/N of the spectrum and the poorly characterized shape of the
underlying continuum. For the pE source CIT 6, the \hal\ line is
strongly blended with other emission features and, therefore, the line
parameters, which are expected to be very unaccurate, have not been
calculated.

\begin{figure}
\plotone{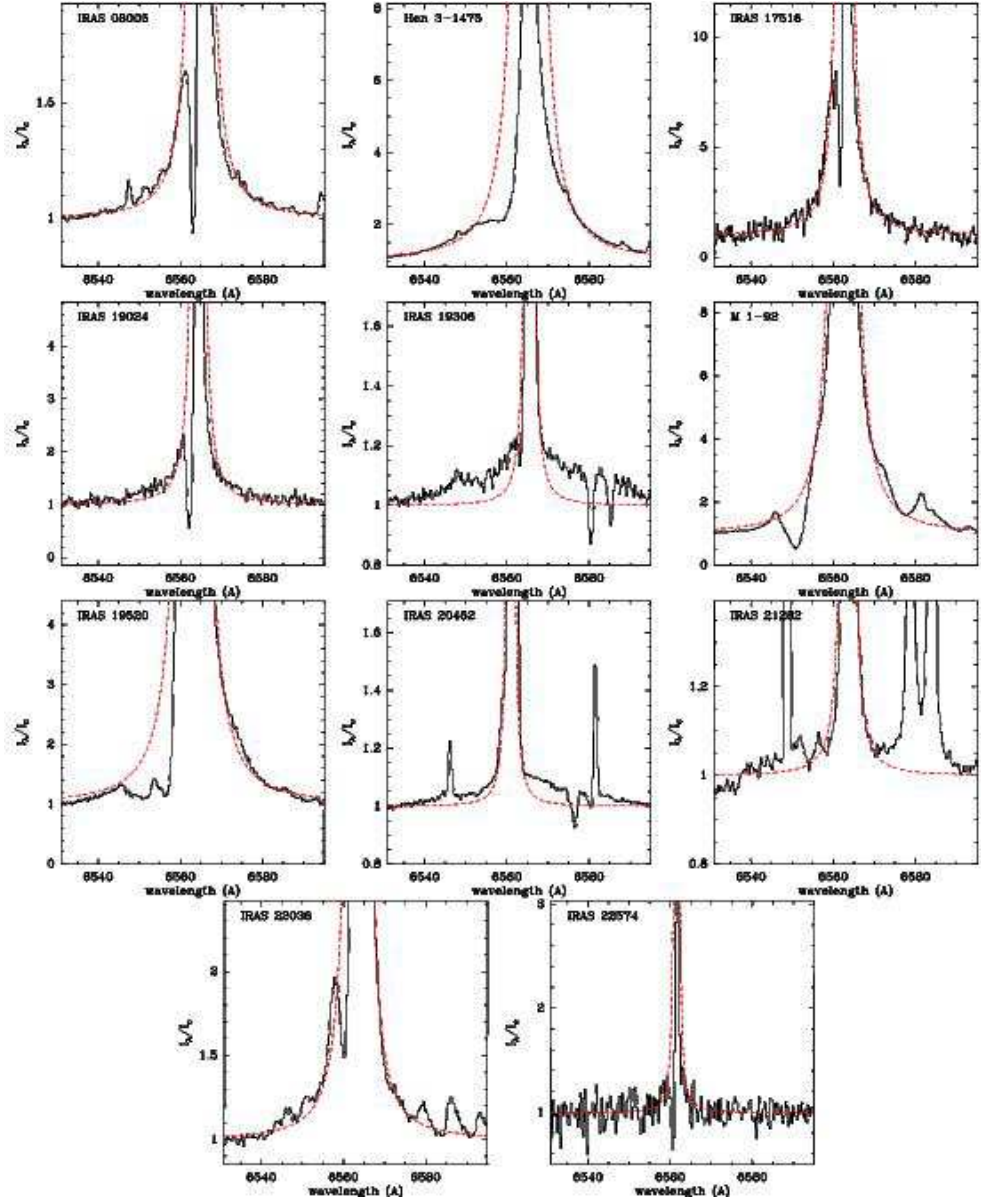}
\caption{Fit to the \hal\ wings of pcyg sources by a function of the type 
  I$_\lambda$\,$\propto$\,$\lambda^{-2}$ (dashed line), which is
  expected for Raman scattering (\S\,\ref{dis_pcyg}). The broad line
  wings of IRAS\,19306$+$1407 and IRAS\,20462$+$3416 cannot be
  reproduced by the fit. \label{f-pcygfit}}
\end{figure}

In order to study whether Raman scattering could explain the broad wings, 
we also fitted the profile of the \hal\ wings with a function of the type
I$_\lambda$\,$\propto$\,$\lambda^{-2}$, which is expected if the wings
are due to Raman scattering (see discussion in \S\,\ref{dis_pcyg}). As
shown in Fig.\,\ref{f-pcygfit}, the fits are satisfactory for all
sources except for IRAS\,19306$+$1407 and IRAS\,20462$+$3416, which
display wings significantly more intense than the synthetic
profile. In these two cases a shallower power law, e.g.\ of the type
$\propto$$\lambda^{-0.7}$, is needed.  We note that IRAS\,19306$+$1407
and IRAS\,20462$+$3416 are among the objects with the broadest wings.

As shown in the table, the FWHM of the \hal\ core emission ranges
between $\sim$50 and 200\kms, whereas the wings reach widths of up to
$\pm$2000\kms. The mean velocity of the bulk of the gas producing the
absorption ranges between $V$$\sim$50 and 500\kms, however, larger
outflow terminal velocities of up to $v_\infty
\approx V_{\rm max} \sim$\,800\kms\ are observed 
(see \S\,\ref{dis_pcyg}). No correlation has been found between the
FWHM of the \hal\ core component and the FWZI or $W_\lambda$ of the
line (Table\,\ref{t_pcyg}). We also investigated the correlation
between the \hal-line parameters and the stellar spectral type but no
obvious trend has been found: although it is true that the largest
equivalent widths ($W_\lambda$$\ga$100\AA) are observed for objects
with hot, O- and B-type central stars, very small values of
$W_\lambda$ ($\la$20\AA) are found for IRAS\,19306$+$1407,
IRAS\,20462$+$3416, and IRAS\,21282$+$5050, which have central stars
of similar early spectral types.  The low values of $W_\lambda$ in
these cases are comparable to those measured in pcyg sources with
stars with types later than B. For these cool (A through G) stars, no
detectable \hal\ emission is expected at all if photoinization were
the only excitation agent in the stellar wind. The relatively
large values of $W_\lambda$ found in these cases are consistent with
shocks being an important ionization mechanism. Alternatively, the
large values of $W_\lambda$ may indicate an additional source of
energetic UV photons (for example a hotter companion?). In a future
work, we will attempt determining quantitatively the importance of
shock excitation in the stellar wind using diagnostic diagrams for
other emission lines arising in the close stellar environment.

The observed values of $W_\lambda$ for objects with the hottest 
central stars in our sample are consistent with UV
stellar radiation being the main ionization mechanism. In this case,
the \hal-line flux is directly proportional to the number of Lyman
continuum photons (N$_{\rm Ly}$) emitted by the star. This is true if the 
\hal-emitting region is radiation bounded.
Under this assumption, we have computed $W_\lambda$ for \hal\ emission
arising in a \ion{H}{2} region around the star using the ionizing
fluxes provided by \cite{vac96} and \cite{smi02} for different
spectral types and adopting a conversion factor from N$_{\rm Ly}$ per
second to \hal\ luminosity of 3.167$\times$10$^{-12}$\,erg
\citep{bro71}. The intensity of the stellar continuum near \hal, which is needed to
calculate $W_\lambda$, has been estimated from the M$_V$ and $R$$-$$V$
values given by \cite{vac96} and
\cite{maiz04}
\citep[see also][]{cox00}.  The flux that corresponds to the zero
$R$-band magnitude, 2.19$\times$10$^{-9}$ erg\,s$^{-1}$\cm2\AA$^{-1}$,
is taken from \cite{fuk95}. The nebular continuum has been also
estimated and taken into account to derive $W_\lambda$. We find that
the apparent $W_\lambda$ dichotomy among pcyg sources with hot central
stars of similar spectral types (with values of
$W_\lambda$$\ga$100\AA\ in some cases and $W_\lambda$$\la$20\AA\ in
others) could easily result from uncertainties in the spectral
classification. 
This is because N$_{\rm Ly}$ and, thus, $W_\lambda$ are very sentive
to the stellar \teff\ for late-O and early-B stars. In particular, for
class I sources with
\teff\ of 28.1 (B\,0), 26.3, and 25\,kK (B\,0.5) we find 
values of $W_\lambda$ of 300, 80, and 50\AA, respectively, i.e$.$\ there
is a factor of 6 in $W_\lambda$ for a difference of only 0.5 spectral
subtypes. (The steep decline of $W_\lambda$ with \teff\ is because the
stellar SEDs peak near the Lyman discontinuity for [O,B]-type stars.)
For O9 stars, we obtain predicted values of $W_\lambda$$\sim$1000\AA,
i.e$.$\,larger than observed ($\sim$20 and $\sim$130\AA,
respectively, for IRAS\,21282$+$5050 and IRAS\,19520$+$2759). In the
case of IRAS\,21282$+$5050, whose central source may be a Wolf-Rayet
[WC11] star, the discrepancy between the observed and expected value
of $W_\lambda$ is smaller since WR stars produce less Lyman continuum
than non-WR stars with the same \teff\ \citep[see][]{smi02}. Values of
$W_\lambda$ smaller than those predicted could also result in the case
of a matter bounded \ion{H}{2} region, for example, a non-spherically symmetric
\hal-emitting region: if the \hal\ emission arises in a
disk or a shell with cavities along a given axis, a large fraction of
the ionizing UV photons will escape along the poles.

\subsubsection{EfA and pA sources}
\label{prof_efA}

\begin{figure}
\plotone{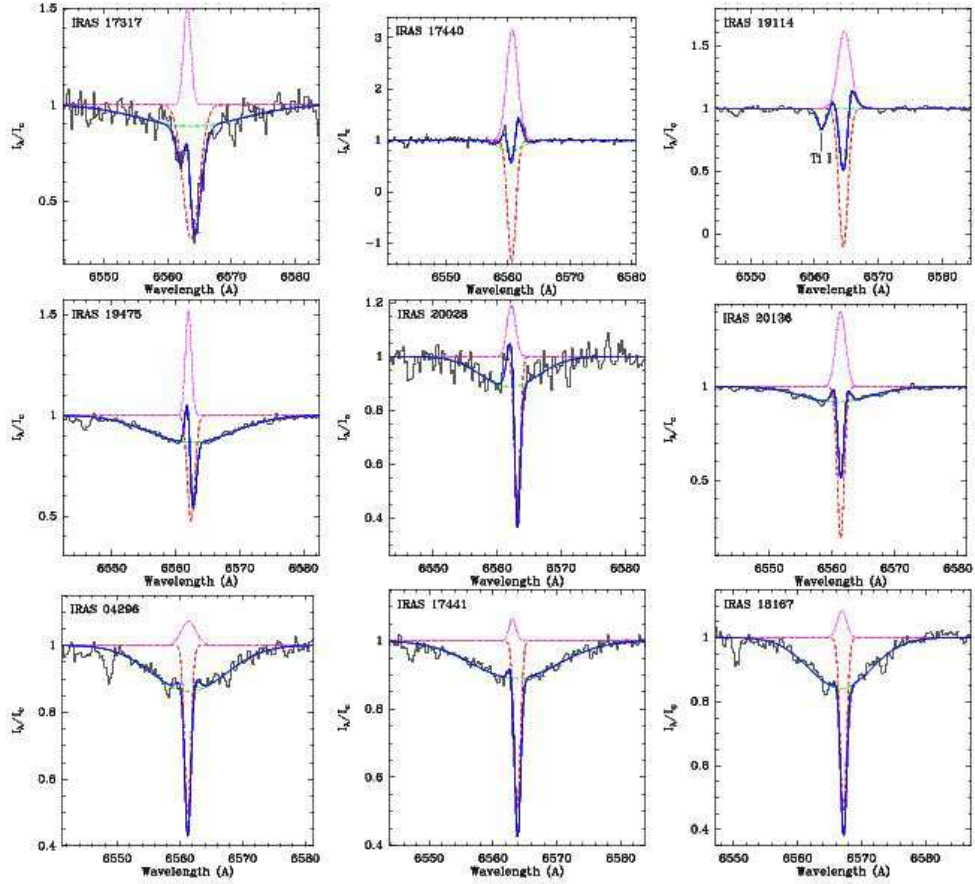}
\caption{Fit to the \hal-profile for efA sources using Gaussian functions ({\sl dotted line}= emission component; {\sl dashed line}= absorption 
  components; {\sl solid line}= resultant profile). Line parameters
  obtained from the fit are given in Table\,\ref{t_gaussfit}.  For
  IRAS\,19114$+$0002, one gaussian function has been fitted to the
  \ion{Ti}{1} line adjacent to \hal\ to better reproduce the observed
  profile. \label{f-gaussf3c}}
\end{figure}

\begin{figure}
\plotone{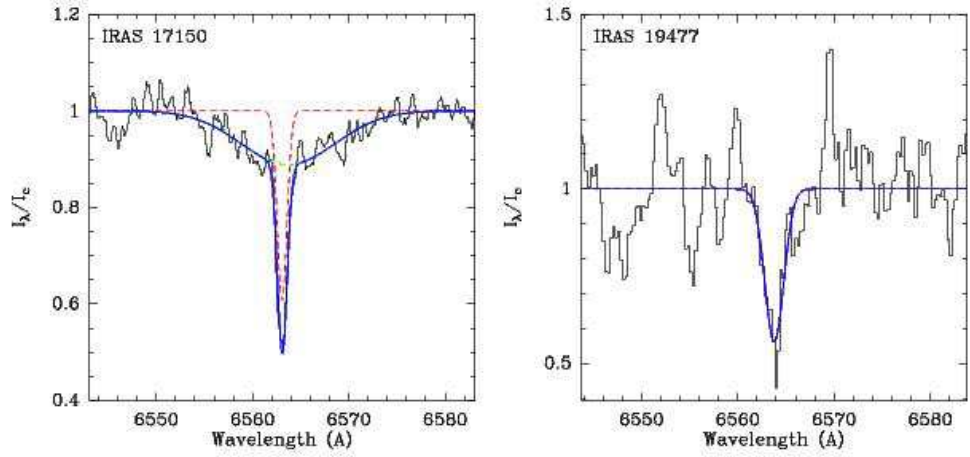}
\caption{Gaussian fit to the \hal-profile for pA sources ({\sl dashed line}= absorption components; {\sl solid line}=
  resultant profile).  
  Parameters of the fit
  are given in Table\,\ref{t_gaussfit}. \label{f-gaussf1c}}
\end{figure}

In order to estimate the relative Doppler shift ($\Delta\lambda$),
equivalent widths ($W_\lambda$), and FWHM of the absorption and
emission components of the \hal\ line for efA+pA sources we have
decomposed the observed profile using different Gaussian functions
(Figs.\,\ref{f-gaussf3c}-\ref{f-gaussf1c} and
Table\,\ref{t_gaussfit}). Two absorption components, one narrow and
one broad, are needed in most cases. 
For IRAS\,19477$+$2401 and IRAS\,19114$+$0002 we have only fitted one
narrow absorption component, however, a weak broad absorption feature
cannot be ruled out: for IRAS\,19477$+$2401 such broad absorption may
be well hidden within the low S/N of our spectrum, and for
IRAS\,19114$+$0002, it could be totally masked out by the relatively
intense emission that fills in the absorption profile.
For IRAS\,19114$+$0002 and IRAS\,17440$-$3310, the line parameters
derived are especially uncertain, in particular, the errors in
$\Delta\lambda$ are probably larger than those quoted in the
table. This is because the absorption features are almost completely
filled in with the emission and, therefore, the fit to the different
absorption components is poorly constrained. This leads to a poor determination of the 
parameters of the emission feature as well. 
Finally, the observed \hal\ profile for the pPNs IRAS\,04296$+$3429,
IRAS\,17441$-$2411, and IRAS\,18167$-$1202, which shows an emission
``hump'' bluewards of the narrow absorption core, can also be
reasonably well fitted without the emission (line parameters with or
without the emission component are given in
Table\,\ref{t_gaussfit}). Therefore, although the line shape is better
reproduced including a weak emission feature, the line parameters
derived for this component are somewhat uncertain.

Although the profile of photospheric absorption lines may be better
represented by a Lorentz or a combined Lorentz-Gaussian (Voigt)
function, we don't expect the parameters derived using these more
complex functions to be significantly different than those in
Table\,\ref{t_gaussfit} given the good match between the Gaussian fit
and the observed line shape. Another way of attempting the
characterization of the emission component in efA sources is to
subtract from the observed profile a synthetic absorption line
obtained from stellar atmospheric modeling. However, for an accurate
characterization of the fundamental stellar parameters that determine
the shape of the absorption profile, such as \teff, gravity,
turbulence velocity, etc, we need to identify and fit simultaneously
not only \hal\ but all the absorption features in our spectra, which
is beyond the scope of this paper.

\subsection{Spectral classification: line absorption spectra}
\label{specclass}

We have used our ESI and MIKE spectra to derive the spectral type of
our targets and investigate the correlation of this fundamental
stellar parameter with the type of \hal\ profile. For some of our
targets, there was already an estimate of the spectral type available
in the literature (see e.g. ``The Torun catalogue of Galactic pAGB and
related objects'' at http://www.ncac.torun.pl/postagb and references
given therein). Whenever no, or uncertain, spectral classification was
found, we estimated the spectral type by comparing the normalized
spectra of our targets with those of template stars obtained from
published stellar libraries.  Note that the slope of the continuum
cannot be safely used for spectral classification due to the uncertain
flux calibration in the majority of the objects in our sample, which
were observed under non-photometric conditions, and also because these
objects are expected to be very reddened by large amounts of dust in
their envelopes.  We have also double-checked the spectral types
previously assigned to our sources. Our template spectra have been
taken mainly from the Library of High-Resolution Spectra of Stars from
the UVES Paranal Observatory Project
\cite[][http://www.sc.eso.org/santiago/uvespop/interface.html]{bag03}.
We have also used as secondary templates some of our own targets for
which a reliable spectral classification was already assigned, e.g.\
from detailed studies of the stellar parameters through
model-atmosphere methods. We have assumed luminosity class I for our
objects. For several sources, this assumption is in fact confirmed
from the strength of the luminosity-dependent \ion{O}{1} infrared triplet (\S\,\ref{lum}).
The main reference used for line identifications and laboratory
wavelengths is the NIST Atomic Spectra Database (Version 3.1.0). In
Table\,\ref{t_parms} spectral types obtained from this work are given
using bold-face font.

In the following we briefly discuss the spectral types assigned in
this work to individual sources with no spectral type available from the 
literature and objects for which our spectral classification does
not agree with previous assignations.

{\it IRAS\,04296$+$3429}. The central star of this object was
classified as a G0\,Ia supergiant based on low resolution optical
spectra by \cite{hriv95}. Detailed chemical analysis and determination
of atmospheric parameters on the basis of high-resolution spectra
performed by different authors, however, indicate a higher effective
temperature of \teff$\sim$6400-7000\,K \citep{klo99,dec98,win00}. By
comparing our high-resolution data with F and G UVES template
standards we find that the spectrum of IRAS\,04296$+$3429 is, in fact,
inconsistent with a G spectral type (e.g., Paschen lines are extremely
weak in G-type stars) but, rather, supports the classification of the
central star as an early-F. In particular, the best match is obtained
with the UVES standard HD 74180 (F3\,Ia).

{\it IRAS\,05506$+$2414}. This enigmatic IRAS source, which is most
likely associated with a YSO, was serendipitously discovered in our
survey of pPNs (¡see \S\,\ref{sample} and Sahai et al., 2008). Its
central source is listed as an M\,6 in the SIMBAD database.  Our ESI
spectrum, however, is not consistent with such a late spectral type
(note, e.g., the remarkable differences with respect to the spectra of
M-type stars in our sample). The deep infrared \ion{Ca}{2} triplet
together with the absence of the Paschen series is indicative of a G-K
star.  Considering also the number and depth of metal lines in the
8400-8800\AA\ window, e.g.\,\ion{Ti}{1} and
\ion{Fe}{1} lines, we assign a spectral type G9-K2 to the central source of this object.

{\it IRAS\,17150$-$3224}. The central star of this object has been
previously classified as G2\,I based on a low-spectral resolution
spectrum in the 4000-6800\AA\ range \citep{hu93}.
However, the density and depth of the lines across the whole
wavelength range observed with MIKE is most consistent with an F3-7
spectral type. (The windows 3890-4600, 5200-5400, 5800-6400, and
7050-7250\AA\ are particularly useful for discriminating between F and
G spectral types.) In particular, the shape of the profile of the
\ion{Ca}{2} H\&K lines (at $\sim$3950\AA) indicates an F3-7 star.  The
spectrum of the template UVES star HD136537, which is a G2 II, has a
much more dense absorption line spectrum and the lines are deeper than
in IRAS\,17150$-$3224.

{\it IRAS\,17317$-$2743}. The central star of this object is
classified as a F5\,I based on a low-resolution optical spectrum in
the 4000-6900\AA\ range by \cite{sua06}. We observe very deep near-IR
\ion{Ca}{2} triplet lines (in the 8500-8700\AA\ range), which is consistent with an F or G spectral
type. However, the absence of Paschen lines series indicates a G-type
star. The spectrum of IRAS\,17317$-$2743 best resembles that of the 
UVES template star HD 204075, which is a G4\,Ib.

{\it IRAS\,17440$-$3310}. There is no previous spectral classification
for this object available in the literature.  The spectrum of
IRAS\,17440$-$3310 is quite similar to that of IRAS\,17150$-$3224
suggesting an F-type star.  In the region of the \ion{Ca}{2} H\&K lines
as well as in the 5200-5400, 5800-6000, and 6100-6300\AA\,
ranges, which are particularly useful to distinguish between early and
late F spectral types, the spectrum of IRAS\,17440$-$3310 is most
consistent with an early-F spectral type. In the following we
adopt an spectral type F3\,I given the good agreement with the spectrum
of our template UVES star HD 74180 (F3\,Ia).

{\it IRAS\,17516$-$2525}. No previous spectral type assignation is
reported for this object. We are unable to obtain an accurate spectral type for this
source since our ESI spectrum does not show any absorption lines but
rather is dominated by nebular emission lines.  The shape of the
spectrum at long wavelengths enables ruling out an M-type
classification. We note the remarkable similarity between the
line-emission dominated spectrum of IRAS\,17516$-$2525 and that of
IRAS\,19520+2759, especially redwards of $\sim$7400\AA. Since the
latter is tentatively classified as an O9 (see below), this may
suggest a hot central star also in this case.


{\it IRAS\,18167$-$1209}. There is no previous spectral classification
for this object available in the literature. Its spectrum
is similar to that of IRAS\,19114$+$0002, suggesting a mid- or
late-F type. The shape of the profiles of the near-infrared
\ion{Ca}{2} triplet and Paschen lines in the range
$\lambda\lambda$8400-8900 as well as the number and depth of the lines
across the wavelength range covered by ESI (but particularly in the
5000-5600 and 6100-6300\AA\ windows) are very similar to those
observed in the UVES template star HD 1089668, F7\,Iab.



{\it IRAS\,19292$+$1806}. No spectral classification has been
previously reported.  The S/N in our spectrum is too low in most of
the wavelength range observed, including the region around \hal.
Relatively weak Paschen lines are seen in absorption in the
8400-8900\AA\ window, which may indicate a B spectral-type in this
case.

{\it IRAS\,19306$+$1407}. Discrepant spectral types have been
previously assigned to the central star of this object by \cite{kel05}
(B0:e) and \cite{sua06} (G5). Both works are based on low-resolution
optical spectra.  A value for the stellar temperature of
\teff=21,000\,K, typical of a B1 type star, has also been found by
\cite{low07} from detailed SED modeling leaving the stellar
temperature as a free parameter. The absence of the near-IR
\ion{Ca}{2} triplet in our ESI spectrum argues against a G- or even an
F-type assignation. Our ESI data support a B0-1 type central star
based on the presence of relatively intense He I lines (e.g.\ at 5876
and 6678\AA), the density and depth of many metal absorption lines,
e.g., in the 4500-4700 and 5600-5800\AA\ ranges, which are typical of
early-B stars, and the relatively weakness of the Paschen lines. The
presence of the
\ion{C}{2} lines at 6578.05 and 6582.88\AA\ in IRAS\,19306$+$1407
further supports its B-type classification, since these lines are very
prominent in B-type stars but extremely weak or absent in other
spectral types. In the following we adopt a spectral type B0-1\,I
given the similarity between IRAS\,19306$+$1407's spectrum and the
template UVES stars HD 112272 (B0.5Ia) and HD 148688
(B1\,Ia). Discrepant spectral types for this target (and other objects) are
discussed in detail in Appendix\,\ref{starevol}.


{\it IRAS\,19374$+$2359}. 
The spectrum of this source is similar to that of IRAS\,19306$+$1407,
suggesting a B-type also in this case. The very weak P$\alpha$ lines,
the absence of metallic lines in the blue, e.g., in the
$\lambda\lambda$5100-5400 region, the presence of
the \ion{He}{1}\,6678\AA\ line, and the very deep
\ion{C}{2}\,$\lambda\lambda$6578,6583\AA\ lines, 
unequivocally indicate an early B-type star. 
The absence of strong metallic lines in the 5600-5700\AA\ range
and the relatively strong \ion{O}{1} triplet suggest a spectral type
later than B2.
The spectrum of IRAS\,19374$+$2359 best resembles that of the UVES
template stars HD 168625 (B2/5\,Ia) and HD 105071 (B6\,Ia/Ib),
therefore, we adopt an intermediate B3-6\,I classification.

{\it IRAS\,19477$+$2401}. \cite{sua06} classify the central star of
this object as an F4-7\,I based on low-resolution optical spectra. We
find, however, that the spectrum of IRAS\,19477$+$2401 is rather
different from those of objects with F3-F7\,I central stars in our
sample (see Table\,\ref{t_parms} and Fig.\,\ref{f_fullspec}). 
The spectral differences are especially noticeable in the 8400-8900\AA\
window, in which the weak Paschen lines and strong near-IR \ion{Ca}{2} triplet
lines are indicative of a later spectral type. The spectrum of
IRAS\,19477$+$2401 is most consistent with the UVES template G0\,Ia
star HD 174383. 

{\it IRAS\,19520$+$2759}. There is no previous spectral classification
for this object in the literature. Spectral typing is difficult in
this case, since the spectrum of IRAS\,19520$+$2759 is dominated by
intense nebular emission lines. The spectrum of this object shows a
general lack of absorption lines, suggesting that it has a fairly hot
star. For example, there is a clear absence of lines in the wavelength
window around 5700\AA. This region shows the presence of numerous
lines for B1\,Ia spectral types (see, e.g., our ESI spectra of
IRAS\,20462$+$3416), but not for hotter spectral types, e.g., O9\,Ia.  The line
absorption spectrum of IRAS\,19520$+$2759 is similar to that of
IRAS\,21282$+$5050, which is classified as an O9.5\,I.  
Therefore, we tentatively assign a similar spectral type, $\sim$O9, to
this source.

{\it IRAS\,20028$+$3910}. The central star of this object has been
previously classified as a G4 \cite[][and references therein]{kel05}.
The presence of the near-IR \ion{Ca}{2} triplet is consistent with
both a G and F type, however, the depth of the Paschen lines indicates
a spectral type earlier than G0. The spectrum of IRAS\,20028$+$3910 is
very similar to that of IRAS\,04296$+$3429, IRAS\,18167$-$1202,
IRAS\,17441$-$2411, which suggests a mid-F classification. In fact,
the relative intensity of the \ion{Ca}{2} triplet and Paschen lines is
most consistent with an F3-7.

{\it IRAS\,20136$+$1309}. The central star of this object has been
previously classified as a G0:I \cite[][and references
therein]{kel05}. The spectrum of this object is similar to that of
IRAS\,20028$+$3910. It shows very deep near-IR \ion{Ca}{2} triplet
lines and relatively intense Paschen lines indicative of a mid- or
late-F spectral type. The number and depth of many metal lines across
the spectrum as well as the shape of the profile of the
\ion{Ca}{2} H\&K lines support an F3-7 assignation. 

{\it IRAS\,22574$+$6609}. There is no previous spectral classification
for this object.  Since IRAS\,22574$+$6609 is very faint in the
optical ($V$$\sim$21.3), our spectrum has a poor S/N at wavelengths
shorter than $\sim$6500\AA\ and no absorption features are identified
in this region.  The infrared window $\sim$8300-8900\AA\ is, however,
very useful for spectral typing:
the depth of the \ion{Ca}{2} triplet lines, which are very weak, and their
relative intensity with respect to the Paschen lines constrain the spectral type to A1-6. The
best-matching UVES template star is HD 80057, which is an A1\,Ia.

\subsection{Luminosity and distance} \label{lum}


The combined strength of the lines at 7771.94, 7774.17, and
7775.39\AA\, of the \ion{O}{1}$\lambda$\,7773\AA\, infrared triplet has been
established as a powerful measure of the stellar luminosity
\citep[e.g.][and refereces therein]{are03}. 
These lines form as a result of upward transitions from the metastable
$3s^5S^0_2$ level, which has a high excitation potetial of 9.15\,eV,
to the $3p^3P_{3,2,1}$ levels of neutral oxygen.  The \ion{O}{1}
infrared triplet is particularly strong in supergiants, especially for
class Ia objects, and is easily observable for stars with spectral
types from middle B to early G. For later G-type stars the \ion{O}{1}
7773\AA\ feature weakens progresively \citep{far88}.

We have used the relationship between the absorption equivalent width
of the triplet, $W_\lambda$(\ion{O}{1}), and the absolute visual
mangitude of the star, $M_V$, calibrated by \cite{slo95}, to derive
the luminosity class of the objects for which the
\ion{O}{1}$\lambda$\,7773\AA\ feature appears in absorption.
The \ion{O}{1} triplet is partially resolved in our observations,
i.e$.$, the 7771.94 line is observed partially separated from the
7774.17 and 7775.39, which appear blended. For IRAS\,08005$-$2356,
Hen3-1475, and IRAS 22036+5306, $W_\lambda$(\ion{O}{1}) cannot be
accurately measured because the \ion{O}{1}$\lambda$\,7773\AA\ feature
shows a P Cygni profile and we cannot rule out some underlying
\ion{O}{1} emission partially filling the absorption.  In addition,
the \ion{O}{1} lines may be blended with some other absorption
features falling in that wavelength range. Therefore, we have not
estimated $M_V$ for these objects. The derived values of $M_V$ (Table
\,\ref{t_oi}) agree quite well (within 0.2$^m$) with the mean of $M_V$
obtained from eqs.\,[1] and [2] by \cite{are03}. (These authors obtain
two separate relationships between $M_V$ and the equivalent width of
the \ion{O}{1}\,7771 line, eq.\,[1], and the \ion{O}{1}\,7774+7775
blend, eq.\,[2].). This agreement indicates that the
$M_V$-$W_\lambda$(\ion{O}{1}) relationship used is valid for the
domains A1-G8 and $M_V$ from $-$9.5 to $+$0.35, i.e, for the spectral
and $M_V$ domain covered by most of our targets. One exception may be
IRAS\,19114$+$0002, with a very large value of $M_V$ consistent with
this object being a luminous, massive star. We note, however, that the
relative atmospheric oxygen abundance of its central star has been
found to be 0.5 dex higher than that in normal F supergiants
\citep{zac96,rh99}, which could result in an overestimate of the
luminosity as derived from the $M_V$-$W_\lambda$(\ion{O}{1})
relationship. The $M_V$-$W_\lambda$(\ion{O}{1}) relationship may not
provide reliable estimates of the luminosity either for the yPNs IRAS\,19306$+$1407,
IRAS\,19374$+$2359, and IRAS\,22574$+$6609, which have B-type central
stars.

The errors of $M_V$ are expected to range between $\sim$0.7 and
1.2$^m$, taking into account the errors in the coefficients of the
$M_V$-$W_\lambda$(\ion{O}{1}) relationship and 
the error of our measurement of $W_\lambda$(\ion{O}{1}), which is expected to
be $\lesssim$0.1\AA.  As shown in Table\,\ref{t_oi} most objects have
$W_\lambda$(\ion{O}{1})$>$1\AA. These high values are restricted to Class I
sources \citep{tho79}, implying luminosities of
$\approx$10$^4$-10$^5$\ls. 
In particular, according to these authors (see their Fig.\,1) and
taking into account the spectral type of our targets
(Table\,\ref{t_parms}), the large majority of our sources belong to
luminosity class Ia. IRAS\,20136$+$1309 and IRAS\,19477$+$2401 are the
two only objects that are most consistent with class Ib. We also find
that the value of $W_\lambda$(\ion{O}{1}) measured for
IRAS\,19306$+$1407 is comparable with, but slightly larger than, the
expected value for a B0-1\,Ia star.  We have estimated the total
luminosity ($L_{\rm bol}$) for objects in Table\,\ref{t_oi} using the
values of the bolometric correction (B.C.) that correspond to their
stellar spectral types.

\begin{figure}
\plotone{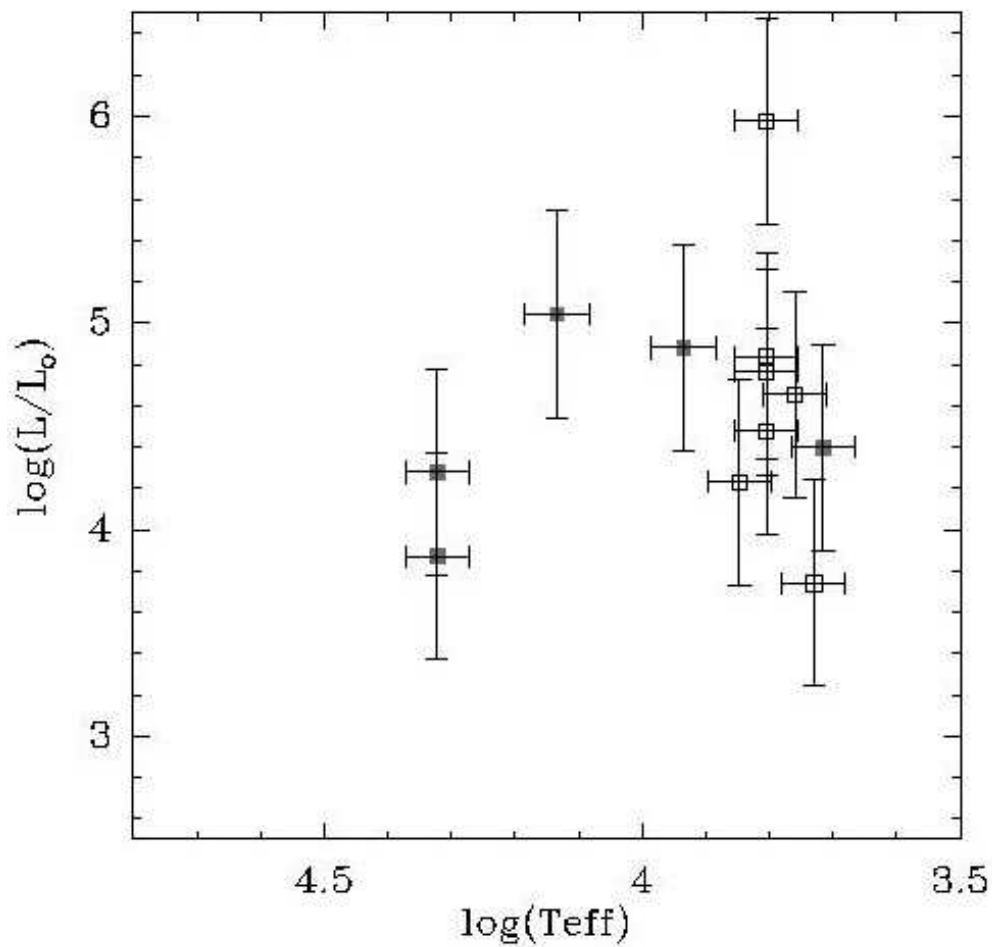}  
\caption{Positions of our targets with the infrarred \ion{O}{1}
  triplet seen in absorption (Table\,\ref{t_oi}), in the HR diagram
  (filled squares: pcyg+pE; open squares: efA+pA). Errorbars of
  $\Delta$log(\teff)=0.05 and $\Delta$log($L$/\ls)=0.5 have been
  adopted. The luminosities derived for objects with B-type central
  stars and for the YHG IRAS\,19114$+$0002 are particularly uncertain
  since these objects could be outside of the applicability domain of
  the $M_V$-$W_\lambda$(\ion{O}{1}) relationship (see text \S\,\ref{lum}).\label{hr}}
\end{figure}

By comparing the position of our targets in the HR diagram
(Fig.\,\ref{hr}) with theoretical evolutionary tracks for pAGB
objects, we find that a significant fraction of our targets (10 out of
13 with $W_\lambda$(\ion{O}{1}) measurements) seem to have central
stars with high remnant masses of $>$\,0.8\msun, which would point
to initial masses of $>$3\msun\
\citep{blo95}. IRAS\,20462+3416, IRAS\,20136+13090 and
IRAS\,19477+2401, are most consistent with lower remnant masses of
$\sim$\,0.6\msun, implying initial masses of $\sim$3\msun.
Note, however, that the large errorbars in the luminosity, $\Delta
log(L_{\rm bol}/\ls)$$\sim$0.5, together with the 
uncertainties arising in the evolutionary models themselves and
the initial-to-final mass relationship, 
prevent us from obtaining accurate values for the masses.

\begin{figure}
\epsscale{0.8}
\plotone{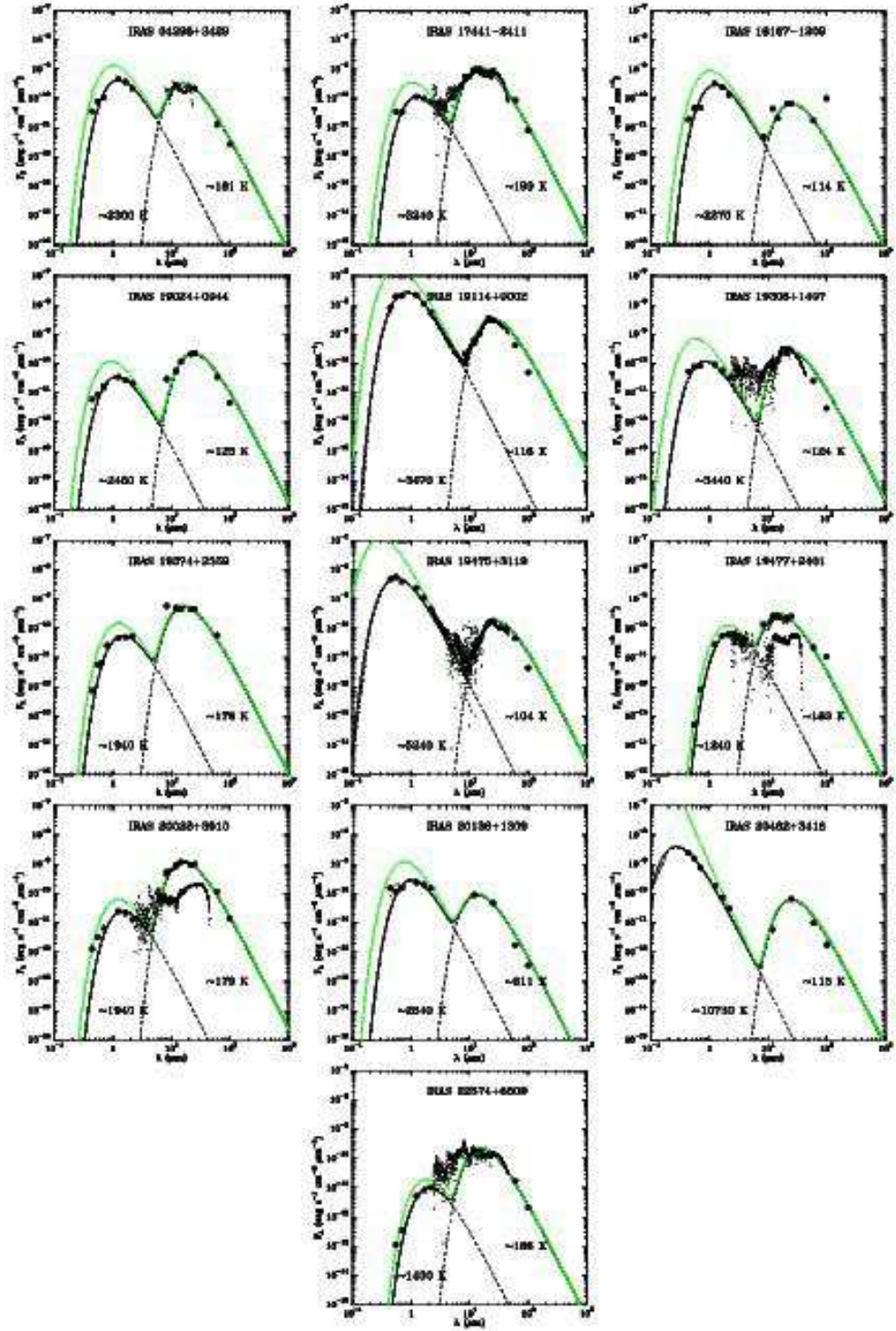}
\caption{SEDs of the objects with infrared \ion{O}{1} triplet
  absorption (Table\,\ref{t_oi}).  Big circles represent GSC2, 2MASS,
  MSX and IRAS data points; Small circles are used for ISO and IRAS
  LSR spectra. Two blackbody curves (dashed lines) have been fitted to
  the reddened stellar photosphere and cool dust emission components.
  The two-blackbody SED model corrected for intestellar extinction is
  also shown (green solid line); an typical value of $A_V$=2.5\,mag is
  used for reddening correction.
\label{bbplots}} 
\end{figure}

We have computed the luminosity of our objects, $L/D^2$, by
integrating their SEDs. We have fitted the double-peaked SEDs by two
blackbody curves, one representing the reddened stellar photosphere,
which is mainly visible from the optical to the NIR, and the other
representing the emission at longer wavelengths by cool circumstellar
dust (Fig.\,\ref{bbplots}). The sum of the integrated fluxes of the
two blackbodies represents $L/D^2$. By comparing the values of $L/D^2$
with $L_{\rm bol}$ we have obtained upper limits to the distance, $D$,
to our objects (Table\,\ref{t_oi}). The upper limits arise because the
total luminosity $L/D^2$ derived from the observed SED is probably
underestimated. This is mainly because our SEDs are not corrected by
interstellar extinction, which can be quite large given the proximity
of our targets to the galactic plane (see below). Correcting for
interstellar reddening, however, is not straightforward since the
total amount of extinction along the line of sight towards our sources
depends on their unknown distances.
Moreover, in some objects there is a clear excess of emission in the
2-10\micron\ range, that is, between the star and cool dust blackbody
curves, denoting the presence of warm dust (see e.g.,
IRAS\,17441$-$2411, IRAS\,19306$+$1407, and IRAS\,22574+6609;
Fig.\,\ref{bbplots}). Since this component is not included in our
simplified two-blackbody model, the value of $L/D^2$ derived from the
model SED is a lower limit to the total luminosity.

We show now that the interstellar extinction correction factor can be
significant in some cases, even if the total extinction is not
extremely high (typical values of the interstellar extinction range
between $A_V$=1 and 4\,mag).  The reddening correction is particularly
important for objects with the stellar component of the SED peaking
shortwards of $\sim$1\micron.  To illustrate this, we have corrected
the SED of IRAS\,20462+3416 for a value of $A_V$$\sim$1\,mag and
adopting a $\lambda^{-1}$ power law for the extinction curve. The
extinction-corrected value of $L/D^2$ is almost a factor 3 larger than
that in Table\,\ref{t_oi}, which translates into a non-negligible
factor of 1/1.7 for the distance. The effect of interstellar
reddening is also exemplified in Fig.\,\ref{bbplots}, were SEDs
corrected by an average value of $A_V$=2.5\,mag are shown.

Finally, there are additional sources of errors that may be affecting
our values of $D$ in Table\,\ref{t_oi} such as the typical
uncertainties in the spectral classification of the central star (of a
few sub-types), which affects B.C$.$ and $L_{\rm bol}$, and the fact
that the reliability of the $M_V$-$W_\lambda$(\ion{O}{1}) relationship
is not well tested for highly evolved objects, specially C-rich
stars, in which the Oxygen abundance may differ significantly from
solar.  Also, for objects with an equatorially-enhanced density
distribution, very common among pPNs, the luminosity $L/D^2$ computed
from the integrated SED depends on the viewing angle, especially when
a significant fraction of the light is scattered by the nebular dust
\citep[see e.g.][for a quatitative estimate of viewing angle effects
on the SEDs of non-spherically symmetric nebulae.]{su01}. High
polarization values, implying mostly scattered light, are indeed
observed for some objects in our sample \citep[e.g.][and references
therein]{tram94,gled05,opp05,bie06}.






\subsection{Correlations} 
\label{corr}

Correlations between the type of H$\alpha$ profile and other stellar
and envelope parameters have been investigated.

\begin{figure}
\plotone{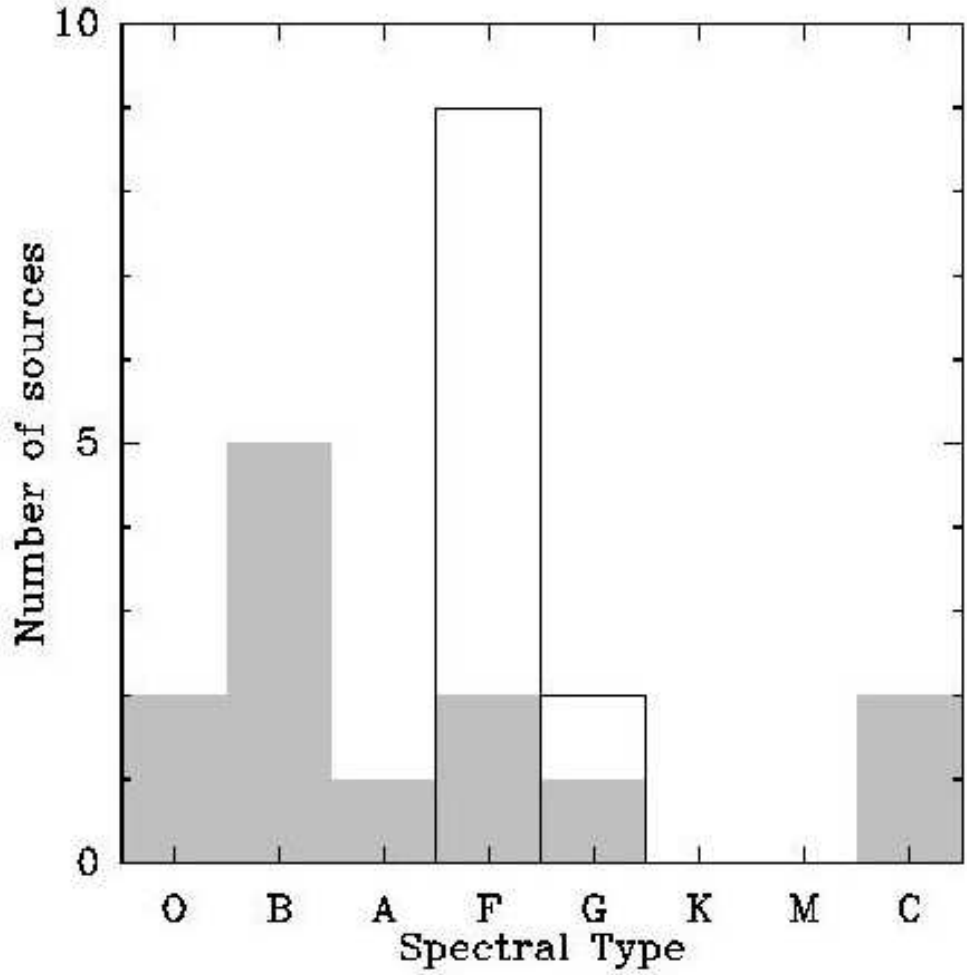} 
\caption{Histogram showing the distribution of spectral 
  types for pcyg+pE (grey-filled bars) and efA+pA (blank bars) sources. \label{f_spt}}
\end{figure}

{\it Spectral type of the central star}. We have found a noticeable,
and somewhat expected, correlation between the \hal\ profile and the
stellar spectral type. First, there is a deficiency of \hal\ emitters
among AGB stars. 3 out of 4 of the \hal\ non-detections are AGB stars
with M or C spectral types.  There are only two late-type stars in our
sample showing \hal\ emission, namely, IRC+10216 and CIT\,6. In these
cases, the \hal\ emission is quite weak ($W_{\lambda}\lesssim$4\AA)
compared with that observed in the majority of the \hal\ emitters
(Table\,\ref{t_pcyg}). This weak \hal\ emission is most likely
temporary, since it was absent in the spectrum of these cool objects
in earlier observations (\S\,\ref{var_hal}).
Second, as shown in Table\,\ref{t_parms} and Fig.\,\ref{f_spt}, all
objects in which \hal\ is seen in absorption, i.e$.$\ efA+pA sources,
have late-type (F-G) central stars.  In constrast, the distribution of
spectral types for
\hal\ emitters, i.e$.$\ pcyg+pE sources, is much broader, including
objects with spectral types ranging from O to G, with a relative
maximum around B, and also very late types such as C. \hal\ emission
is indeed expected in objects with [O,B]-type central stars, whose UV
radiation field ionizes the circumstellar material, producing an
\ion{H}{2} region observable through Balmer line emission
(\S\,\ref{prof_efA}). For objects with mid or late spectral types,
lacking in photoinizing photons, a likely mechanism for
producing \hal\ emission are shocks.  For non-pulsating stars, such
shocks probably result from the interaction between the inner layers
of the AGB envelope, i.e$.$\ the slow AGB wind, and the fast pAGB
winds, more recently ejected. In the two long-period variable stars
with \hal\ emission in our sample, shocks could either arise in the
pulsating stellar atmosphere, which is the normal interpretation for
transient line emission in this type of objects \cite[e.g.][and
references therein]{cas00}, or result from the AGB-to-pAGB wind
interaction. We note that, in fact, such interaction must have started
in these objects given the non-spherical morphology of their
inner CSEs.

\begin{figure}
\plotone{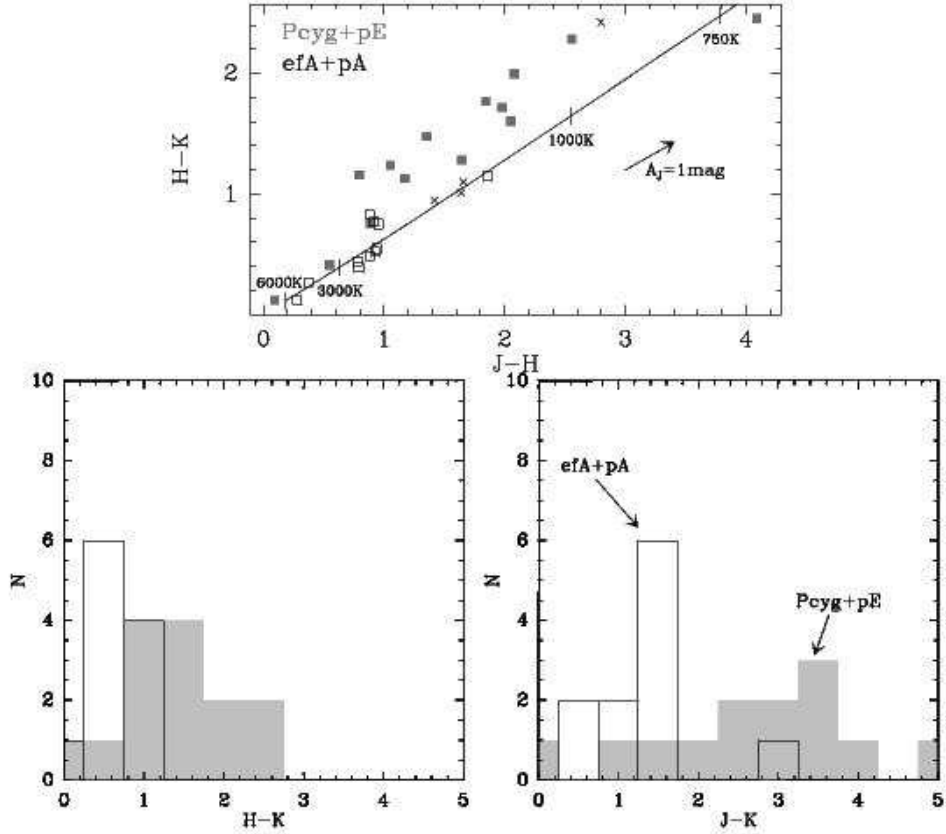}
\caption{($Top$) Distribution of our targets in the ($J-H$)-($H-K$)
  color-color diagram (filled squares: pcyg+pE; open squares: efA+pA;
  crosses: non-\hal\ detections). The reddening vector for
  $A_J$=1\,mag is indicated.  The solid line represents the locus of
  blackbody emitters for temperatures in the range 6000-750\,K; the
  tick marks correspond to the temperatures indicated.  ($Bottom$)
  Histograms of the $J-K$ and $H-K$ colors for pcyg+pE and efA+pA
  sources (same color code as in Fig.\,\ref{f_spt}).
\label{histo_nir}}
\end{figure}

\begin{figure}
\plotone{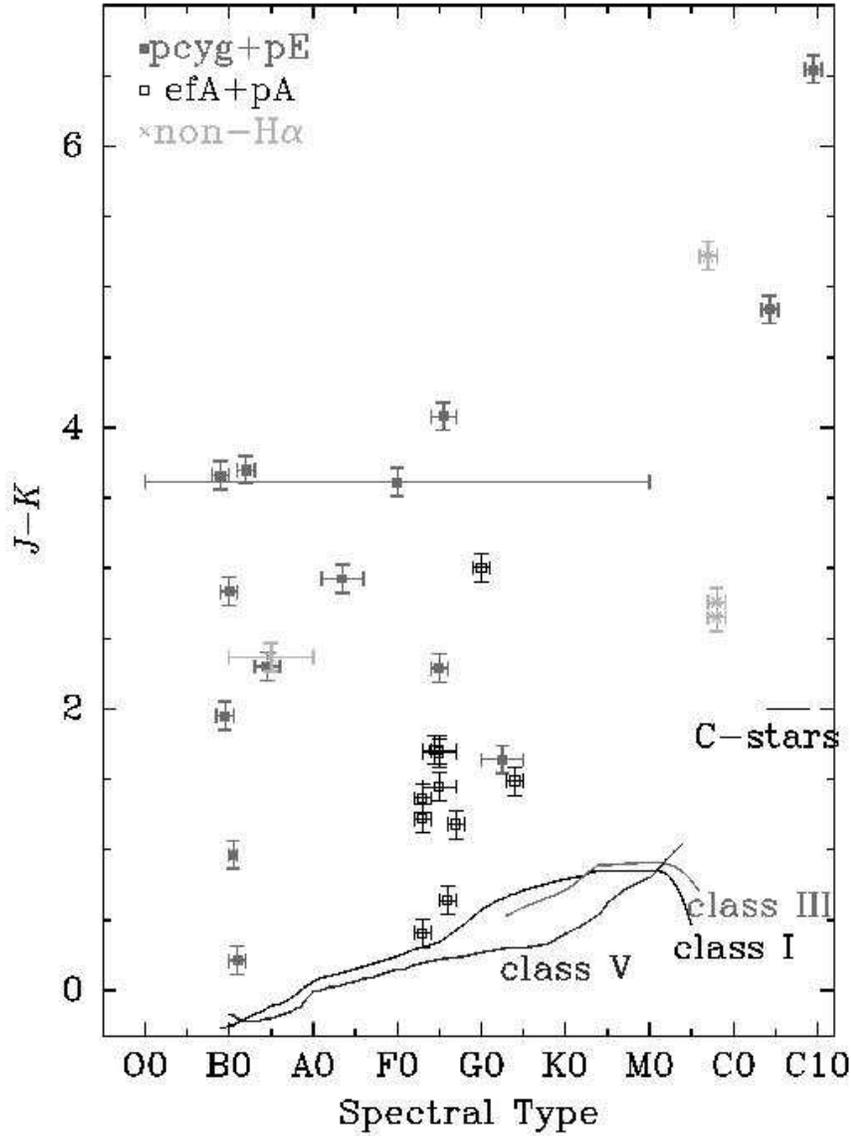}  
\caption{$J-K$ vs$.$ spectral type diagram showing the
  observed colors for our targets (symbols) and the intrinsic colors
  for luminosity classes I, III, and V, and for C-rich stars (solid
  lines; from \cite{duca01}). Horizontal errorbars represent
  uncertainties in the spectral classification of the central stars.
  For the $J-K$ color, an error of 0.1mag has been adopted. Symbol code as in
  previous figures. 
\label{jk_spt}}
\end{figure}

{\it NIR and IRAS colors}. NIR colors are sensitive to the presence of
warm ($\sim$500-1000\,K) dust.  The ($J$$-$$H$)-($H$$-$$K$)
color-color diagram shows that pcyg+pE objects are redder than efA+pA
sources (Fig.\,\ref{histo_nir}). This is also apparent in the
histograms of any of the three NIR colors, especially $H$$-$$K$ and
$J-K$ also shown in the figure. We have compared the observed values
of $(J-K)$ with the intrinsic color ($J-K$)$_0$ for the spectral types
assigned to our targets and found that the larger values of $(J-K)$ in
\hal\ emitters do not result from their different distribution over
spectral type with respect to efA+pA sources (Fig.\,\ref{jk_spt}). The
larger $(J-K)$ color excess in pcyg+pE sources may be partially
explained by a larger value of the total (interstellar+circumstellar)
extinction. However, reddening $only$ cannot explain the loci of most
of these objects above the black-body line in the ($J-H$)-($H-K$)
color-color diagram given the direction of the reddening vector
\citep[from][equivalent to a $\propto$$\lambda^{-1.6}$ extinction
power law]{schl98}. Scattering by nebular dust could also affect the
observed colors. However, since the scattering efficiency is larger at
shorter wavelengths, the scattering net effect is blueing of the
incident light, and that will not explain the observed $H-K$ colors of
pcyg+pE sources, which are redder than for a black body. 
While the NIR colors of efA+pA sources are consistent with the major
contribution to the emission in the NIR bands being the reddened
stellar photosphere, the position of pcyg+pE sources in the
($J-H$)-($H-K$) color-color diagram highlights the presence of warm
dust. This warm dust component is in fact confirmed in some cases by a
clear emission excess between 2 and 10\micron, whenever such data
exists (see SEDs of, e.g. IRAS\,19306$+$1407 and IRAS\,22574$+$6609 in
Fig.\,\ref{bbplots}). The warm dust component must be near the star
and, therefore, has probably been formed recently. Then, the presence
of warm dust could be an independent confirmation of substantial
present-day mass-loss in pcyg+pE targets, which is evidenced in the
first place by their intense, circumnuclear \hal\ emission.
Alternatively, the warm dust component in pcyg+pE sources could be
located in a long-lived disk around the central source rather than in
the pAGB wind itself.  Finally, we note that the NIR colors of the efA
objects IRAS\,17150$-$3224, IRAS\,17441$-$2411, and IRAS\,20136$+$1309
also suggests dust grains with higher temperatures than those in their
detached, much cooler, AGB CSEs (see also full SEDs for the two later
in Fig.\,\ref{bbplots}).

We have compared the location of our objects with the larger,
un-biased sample used by \cite{gl97b}. Our pcyg sources fall in
regions III and IV defined by these authors, which suggest that the
NIR colors of objects in such regions may result from recent mass loss
leading to a hot dust component around the star. For a few of their
objects for which optical spectroscopy was available, P-cygni \hal\
profiles were found. Our work has demonstrated that pcyg sources
systematically lie above the black-body line in the NIR two-color
diagram, displaying colors that are consistent with the presence of
hot dust in the stellar vicinity most likely resulted from pAGB
mass-loss. There are two pcyg objects, IRAS\,20462$+$3416 and
IRAS\,19306$+$1407, with black body-like NIR colors
(Fig.\,\ref{histo_nir}) that can be explained just by reddening of the
stellar atmosphere. For these two pcyg objects, therefore, there are
no indications of the presence of hot dust in the stellar vicinity
although their P-cygni \hal\ profile indicates current mass-loss. In
IRAS\,20462$+$3416, significant spectral variations with time in the
\hal\ line have been reported, pointing to episodic mass-loss events
(\S\,\ref{var_hal}). As suggested by \cite{gl97} the absence of
significant NIR excess in this case could indicate that the duration
and intensity of such mass-loss episodes are not enough to efficiently
form and maintain hot dust grains in the circumstellar envelope.

\begin{figure}
\plotone{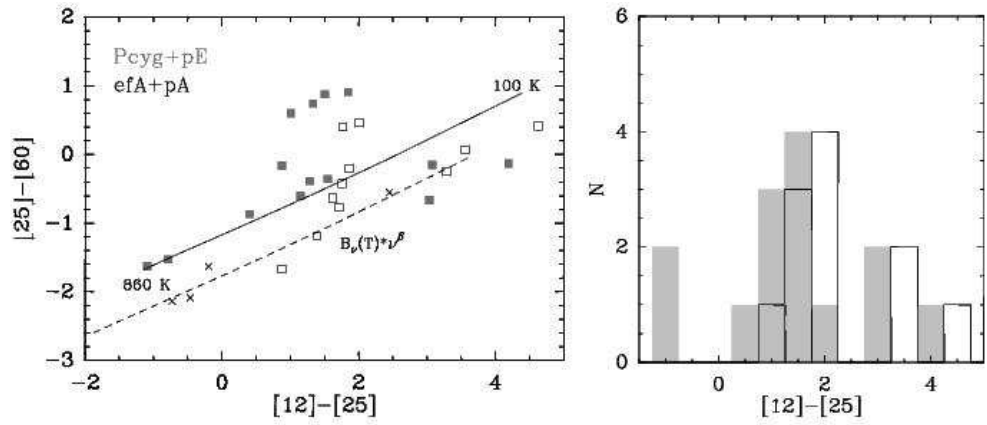}   
\caption{($Left$) Distribution of our targets in the IRAS [12]$-$[25]
  vs$.$ [25]$-$[60] color diagram. The solid line represents the
  blackbody colors for temperatures in the range 100-860\,K. The
  dashed line represents the colors for optically thin dust emission
  for the same range of temperatures and dust optical depth
  $\tau_\nu\propto\nu^{\beta}$ (see \S\,\ref{corr}).  ($Right$)
  [12]$-$[25] histogram --- same color-symbol code as in
  Fig.\,\ref{histo_nir}.
\label{iras_segr}}
\end{figure}

We have also found a correlation between the type of \hal\ profile
and the IRAS colors, which are most sensitive to the cool
($\sim$100-200\,K) dust component that is remnant of the mass loss on
the AGB (Fig.\,\ref{iras_segr}).  The IRAS colors represented in this
figure are defined as [12]$-$[25]=2.5$log$(F$_{25}$/F$_{12}$) and
[25]$-$[60]=2.5$log$(F$_{60}$/F$_{25}$) with F flux density in Jy at
the 12, 25, and 60\micron\ IRAS filters. As can be seen, efA+pA
sources display slightly larger values of [12]$-$[25], i.e$.$ the
slope of the SED between 12 and 25\micron\ is steeper, than for
pcyg+pE. This is consistent with a well detached envelope 
lacking significant amounts of warm dust in most efA+pA sources, which is
indicated as well by their NIR colors (see above). The two groups of
\hal\ profile sources clearly segregate in the two-color IRAS diagram: the
majority of the pcyg+pE sources ($\sim$64\%) lie above the black-body
line, whereas most efA+pA targets ($\sim$73\%) are below. This
separation suggests different properties of the circumstellar dust in
the AGB CSE among the two groups. The distribution of the sources
below the black body line, which means that the emission at the longer
wavelengths is smaller than expected for a black-body, can be
explained by the dependence on wavelength of the dust optical depth,
$\tau_\nu\propto\nu^{\beta}$ (normally $\beta$$\approx$1-2) and the
fact that the dust emission is proportional to $B_\nu$($T_{\rm
d}$)$\times$($1-e^{-\tau_\nu}$), with $B_\nu$($T_{\rm d}$) being the
Plank function for a given dust temperature $T_{\rm d}$.  For
optically thin dust emission the flux is, therefore,
proportional to $B_\nu$($T_{\rm d}$)$\times$$\nu^\beta$ (this is the
so called ``modified'' black-body function), which explains the
smaller emission at the longer wavelengths (Fig.\,\ref{iras_segr}).
The location of the targets above the black body line in the IRAS
color-color diagram means, as for the NIR, that their colors cannot be
well represented by a black body or modified black body with a
$single$ temperature. Such location must result from a larger
temperature distribution of the dust over the envelope in pcyg+pE than in
efA+pA sources. In particular, the observed effect suggests a significant
contribution to the total emission by cold dust, especially at longer
wavelengths (60\micron), whereas most of the emission at shorter
wavelengths (12\micron) would be produced by a different component of
warmer dust grains.

\begin{figure}
\plotone{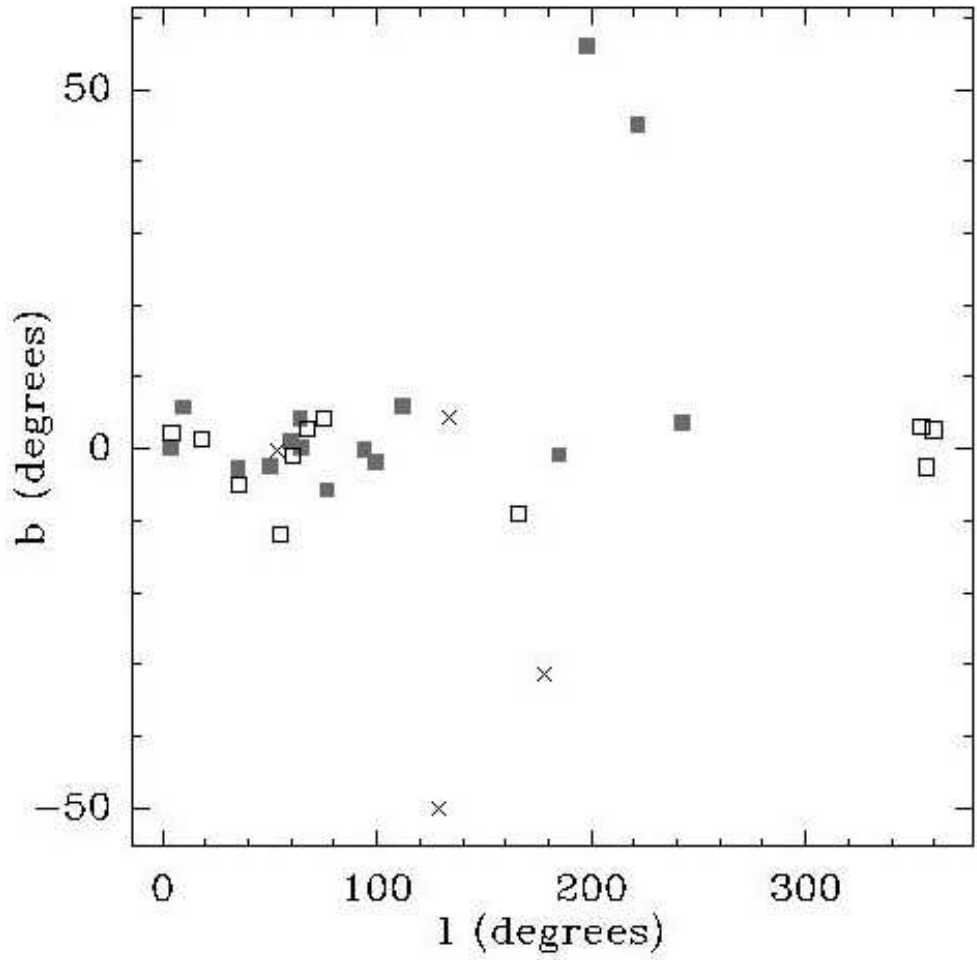}  
\caption{Distribution of our targets in the Galaxy. Symbol code as in 
  previous figures. \label{f_gal} }
\end{figure}

{\it Galactic coordinates.} We have not found any obvious correlation
between the \hal\ profile and the galactic coordinates of our targets
(Fig.\,\ref{f_gal}). A correlation with the galactic latitude, $b$,
would have been proof of differences in the masses of the
progenitor of both types of objects.  We note, however, that most of
our targets happen to be concentrated in a relatively small range of
latitudes around $b$=0\degr, so the dependence of \hal\ with $b$
cannot be properly studied. (The narrow range in $b$ in our sample
results from the fact that OH/IR surveys, from which most our targets
were selected, normally concentrate around the galactic plane.)

{\it Nebular morphology.} We find a variety of nebular morphologies in
our sample but no obvious segregation with respect to the \hal\
profile has been found. There is nearly the same fraction of E, M, B,
I, and stellar sources as well as w- and $\star$ objects among \hal\
emitters and efA+pA targets (see Table\,\ref{t_parms} and references
therein).

{\it Chemistry.} It seems that there is little connection, if any,
between the chemistry and the characteristic of the \hal\
profile. There are more or less the same fraction of O- and C-rich
objects among the \hal\ emitters and efA+pA sources. Among AGB stars,
we find that the only objects with \hal\ emission are C-rich.  Because
of the small number of AGB stars in our sample and the expected
variability of the \hal\ emission with the light curve, this trend
cannot be regarded as a robust result.


\section{Discussion}
\label{discuss}

\subsection{Fast pAGB winds probed by P Cygni like profiles} 
\label{dis_pcyg}

P Cygni profiles are known to be indicative of on-going mass-loss in
the form of a stellar wind.
The formation of typical P Cygni profiles can be understood
qualitatively by a simple model of a spherically symmetric wind in
which the velocity increases outwards up to a terminal velocity of
the outflow $v_\infty$.  The wind material in front of the star is
moving towards the observer and, therefore, absorbs the stellar
continuum producing a blue-shifted absorption feature with a Doppler
shift between $-$$v_\infty$ and 0, relative to the source systemic
velocity (\vsys). Except for the part that is occulted by the star,
the rest of the wind, which forms a halo around the star, will produce
broad line emission centered around \vsys\ with wings extending up to
$\pm$$v_\infty$.  The sum of the emission and the absorption 
at each Doppler velocity range leads to a so called
P Cygni profile. 
These P Cygni profiles are often seen in
UV resonance lines of the central stars of PNs \citep[e.g.][]{peri93,kud97}.

The \hal\ profiles observed in the majority of our targets differ from
the typical P Cygni profiles described above. The emission component
is not centered at \vsys\ but, rather is red-shifted in all cases,
except for IRAS\,21282$+$5050 (Table \ref{t_pcyg}).  Moreover, the
terminal velocity of the outflow, as measured from the blue-shifted
edge of the absorption feature, is smaller than that obtained from the
edge of the red emission wing, and in some cases, e.g.\ the yPN
M\,1-92 (see below), the blue absorption is well detached from the line emission core.
These peculiar H$\alpha$ line-shapes can be explained if the emission
and absorption form in two distinct nebular components, with different
velocity fields and, more generally, affected by different broadening
mechanism. In particular, we propose the following model: the broad
line emission, characterized by a given $intrinsic$ (not necessarily P
Cygni) profile, arises from a compact central source; the line
emission (and stellar continuum) is scattered by dust in the walls of
the nebular lobes, which should produce an overall red-shift of the
line profile as a whole relative to \vsys\ as observed. The
blue-shifted absorption is due to neutral or partially ionized
outflowing gas inside the lobes absorbing the scattered photons along
the line of sight. This results in a P Cygni like profile where the
blue-shifted absorption is produced against the intrinsic emission
profile. Such a model has been successful in producing a detailed fit
to the spatio-kinematic distribution of the blue-shifted absorption
and the red-shifted scattered line core in the H$\alpha$ line profiles
observed with STIS/$HST$ in the yPN Hen\,3-1475 (S\'anchez Contreras
\& Sahai 2001; see their Figure\,1d for a sketch of our geometrical
model). A similar scenario is proposed by \cite{arr05} to explain the
P Cygni like profile of \hal\ and other recombination lines in the yPN M\,1-92,
which shows a very broad, blue absorption feature well detached from
the emission component that cannot result from the single stellar wind
scenario used to explain typical P Cygni profiles.
Our interpretation is also consistent with the fact that most pPNs
exhibit intrinsic polarization of the continuum as well as some
emission lines, which are most likely due to scattering off dust
grains \citep{tram94}. In particular, Balmer lines seem to come from
an \ion{H}{2} region close to the star, which is blocked from direct
view by, e.g., the thick equatorial dust waists commonly present in
these objects, and are seen in reflection off the nebular dust
\cite[see also, e.g.,][]{san02}.


The nature of the outflow producing the absorption in pcyg sources is
unknown, however, as in the case of Hen\,3-1475, we argue that it is
most likely a ``pristine" pAGB wind from (or near) the central
star that has not been strongly altered by its interaction with the
progenitor AGB wind. The projected speed of the bulk of the material
in the pAGB outflow, derived from the centroid of the blue-shifted
absorption feature relative to the systemic velocity in our sample, is
approximately $\sim$100\kms\ (Table\,\ref{t_pcyg}); however the
absorption features are quite wide and reach larger velocities at the
edge indicating the presence of smaller amounts of material expanding
as fast as, e.g., $v_\infty$\,$\sim$800\kms\ in the case of
IRAS\,19520$+$2759.
As for Hen\,3-1475, it is plausible that a radial and or latitudinal
velocity gradient exists in the outflows and/or that distinct wind
components are present; the latter is suggested by the two different
absorption features observed in the \hal\ profiles of a few objects
(\S\,\ref{res_hi}). Unfortunately, ground-based observations do not
allow us to conclude on the isotropic or collimated nature of the
pAGB winds probed by the observed P Cygni profiles, however, it is
plausible that these are bipolar/multipolar ejections, as in
Hen\,3-1475 and, probably, M\,1-92.


\subsection{Broad \hal\ emission wings} 
\label{dis_wings}

The emission component observed in the \hal\ profile in our pcyg
sources consists of an intense core plus weaker broad wings that reach
widths (measured as the line FWZI) of up to $\sim$4000\kms. The yet
unclear origin of the very extended \hal\ wings, which are also
observed in a number of PNs, has been discussed by \cite{arr03}. These
authors consider several possible mechanisms for line broadening and
conclude that Raman scattering of photons with wavelength close to
Ly$\beta$ by neutral hydrogen is the most probable one in 12 of the 13
objects in their sample.  In this scenario, Ly$\beta$ photons with a
given velocity width are converted to optical photons that fill the
\hal\ wing region.
The conclusion of \cite{arr03} is mainly supported by a) the fit of
the \hal-wing profile to a $\propto \lambda^{-2}$ law; b) the presence
of other Raman features produced in emission; and c) the presence of a
significant neutral hydrogen component.  Criteria a) to c) are to be satisfied
for the line wings to be attributable to Raman scattering.

We have checked whether or not the same conclusion can be drawn from
our sample. As we have shown in Section
\ref{proffit}, the \hal\ wings observed follow a 
$\propto \lambda^{-2}$ law in all cases, except for IRAS\,20462$+$3416
and IRAS\,19306$+$1407. In these two objects, which happen to display
\hal\ wings with similar widths and shapes, Raman scattering can be
ruled out as the main line broadening mechanism. \cite{arr03} found
the same disagreement between the observed and predicted Raman profile
for IRAS\,20462$+$3416, which was also in their sample, concluding
that a strong stellar wind is the most probable mechanism for line
broadening in this case. We have searched for other emission features
produced by Raman scattering in our spectra. In particular, we looked
at the 6830 and 7088\AA\ bands that correspond to the Raman-scattered
\ion{O}{6} $\lambda\lambda$1032,1038 doublet. Only four objects,
IRAS\,08005$-$2356, M\,1-92, Hen\,$-$1475, and IRAS\,22036$+$5306,
show emission around 6830\AA\ with FWZI$\sim$150\kms, and only in two
of them, Hen\,$-$1475 and IRAS\,22036$+$5306, a weak emission feature
is tentatively detected around 7088\AA\ with FWZI$\la$100\kms\
(Fig.\,\ref{f_fullspec}). In contrast, \cite{arr03} find these and
other Raman features with large FWZIs of a
few$\times$10$^2$-10$^3$\,\kms, i.e$.$\ comparable to the width of the
\hal\ line wings, in most of their objects.
 
Since both the $\lambda^{-2}$ profile of the wings and the presence of
a large column density of neutral hydrogen are necessary but not
sufficient conditions for attributing the observed \hal-wings to Raman
scattering, we cannot conclude that in our sample this is the main
mechanism for line broadening except maybe for the four objects listed
above showing other Raman-scattered emission features. Note, however,
that the FWZIs of such features are significantly smaller than those
measured in Arrieta's sample and smaller than the \hal\ wings
themselves.

As suggested by \cite{arr03}, further investigation of the effect of
Raman scattering producing the line wings could be obtained by
measuring the widths of UV lines produced in the same region as the
Ly$\beta$ photons. This is because the width of the scattered \hal\ is
proportional to the initial width of the Ly$\beta$ line, the
Raman-scattering broadening factor being of $\lambda_{\rm
H\alpha}$/$\lambda_{\rm Ly\beta}$=6.4. We searched the IUE archive for
UV spectra of our targets around the line \ion{Si}{3}]$\lambda$1892\AA,
which presumably arises in the same region as
Ly$\beta$. Unfortunately, such data only exist for four objects and
only in one case, Hen\,3-1475, the Si III] line is clearly
detected. The line, which was observed with the low spectral
resolution ($\Delta\lambda$$\sim$6\AA) SWP, is found to be spectrally
unresolved. Nevertheless, for Raman scattering being able to produce
the observed $\sim$4000\kms-wide \hal\ wings in Hen\,3-1475, the
Ly$\beta$-emitting region must be characterized by a typical velocity of 
$\sim$600\kms, i.e$.$\ the Si III] line should be spectrally resolved
with a FWHM of $\sim$7\AA\ as observed with SWP, in constrast to what
is found. Therefore it seems that Raman scattering is not the main
mechanism affecting the profile of the \hal\ wings in this case.

One additional requirement for Raman scattering to be efficient is
that a relatively strong incident Ly$\beta$ flux has to be produced by
the central source. This may be satisfied by 
Arrieta's sample, formed by objects with central stars with
\teff$\geq$20,000\,K, except maybe for the hypergiant IRC$+$10420 with
a stellar spectral type of A5 (\teff$\sim$8500\,K) at the time
Arrieta's data were obtained.\footnote{The spectral type of the
central star of IRC$+$10420 changed from F8 to A5 in 4 years, from
1992 to 1996 \citep{klo97}.} An estimate of the Ly$\beta$ luminosity
expected around the central star of the PN IC\,4997 using the
photoinization code CLOUDY has been done by Lee \& Hyung (2000). In
this case, for a very hot (\teff=60,000\,K) and quite small
($R_*$=\rsun) central star, a compact \ion{H}{2} region of high
density ($\approx$10$^9$-10$^{10}$\cm3) together with a column density
of neutral hydrogen of $N_{\rm HI}$=few$\times$10$^{20}$\,\cm2 in the
scattering region are needed in order to have a sufficient number of
Ly$\beta$ photons and efficient Raman scattering. In our sample, a
significant number of objects showing broad \hal\ wings have
relatively cool central stars, with spectral types F-G
(\teff$\sim$5000-7000\,K) and, therefore, are unlikely to produce
enough UV photons around Ly$\beta$. 
This is in fact supported by their
featureless nebular spectra, which are remarkably different from those
observed in more evolved PNs (\S\,\ref{res_nohi}). 
For the objects in our sample with the hottest
central stars (with late-O and early-B spectral types,
\teff$\sim$20,000-30,000\,K) estimates of the Ly$\beta$ flux will be presented
in future individual papers. We will also investigate whether or not
shocks resulting from the wind interaction process are able to provide
a strong Ly$\beta$ flux.

The presence of fast stellar winds is another plausible mechanism to
explain the broad \hal\ wings observed. As already pointed out by 
\cite{arr03}, this is indeed the most likely mechanism for IRAS\,20462$+$3416, 
for which the \hal\ wing profile was found not to follow a $\propto
\lambda^{-2}$ law. For similar reasons, 
a fast ($\sim$1400\kms) stellar wind is also very likely being
expelled by the central star of IRAS\,19306$+$1407 leading to its
\hal\ wings. In more than one third of the objects in our sample, 
the presence of stellar winds is, in fact, confirmed by P Cygni
profiles in \hal\ and other lines. The question is whether or not
these winds reach much higher velocities (of 1000-2000\kms) in certain
regions that can explain the broad wings observed. In the case of
M\,1-92 and Hen\,3-1475 there is direct evidence for fast winds with
expansion velocities of up to $\sim$800 and 2300\kms, respectively
\citep{arr05,san01}, so it is plausible that these winds are also
responsible (at least partially) for their \hal\ wings. In the rest
of the objects in our sample, the terminal speed of the fast wind
measured from the blue-shifted absorption feature of the
\hal\ P Cygni profile ($v_\infty \approx V_{\rm max}$$\approx$10$^2$\,\kms, 
Table\,\ref{t_pcyg}) is never as large as that implied by the broad
wings if these were to be attributed to fast outflows. However, we
find that $V_{\rm max}$ is normally larger than the FWHM of the \hal\
core indicating that indeed there are winds moving faster than the
region emitting the bulk of the \hal\ line (with an average velocity
of $\sim$120\,\kms; Table\,\ref{t_pcyg}).  This clearly suggests the
presence of velocity gradients in the stellar winds of these
objects. Moreover, we have found a weak correlation between $V_{\rm
max}$ and the FWZI of the broad \hal\ wings: the sources with the
largest values of $V_{\rm max}$, namely, IRAS\,19520$+$2759,
IRAS\,20462$+$3416, M\,1-92, and Hen\,3-1475, exhibit the broadest wings,
reaching widths of $\sim$2600, $\sim$2800, $\sim$3000, and
$\sim$4000\kms, respectively.  (For IRAS\,20462$+$3416 and Hen3-1475, the \hal\
absorption feature is partially masked out by emission, therefore, values of
$V_{\rm max}$=$-$300\kms\ are measured from their H$\beta$ P Cygni
profiles; Fig.\,\ref{f-hb}.)  This could be an indication that broad \hal\ wings are 
partially due to fast stellar winds.



%

\subsection{Incipient mass-loss probed by efA profiles} 
\label{dis_efA}

Structured \hal\ absorption profiles partially filled in with emission
are common in F- and G-type supergiants and are considered as a
possible evidence for mass-loss \citep[e.g.][]{tam93,are85,sow90}.
Although most efA sources in our sample are surrounded by extended
nebulosities (see Table\,\ref{t_parms} and references therein), the
observed structured \hal\ profile, in particular, the weak emission
that fills in the stellar absorption, arises in the bright nebular
nucleus, i.e$.$\ in a compact region very close to the central star, and
not in the extended nebulae. This compact nuclear region is directly
seen but is spatially unresolved in our ESI spectra for
IRAS\,04296$+$3424, IRAS\,17440$-$3310, IRAS\,18167$-$1209,
IRAS\,19114$+$0002, IRAS\,19475$+$3119, and IRAS\,20136$+$1309.
Among these, IRAS\,17440$-$3310 is the only efA source for which our
ESI spectrum also shows weak \hal\ emission arising in the extended
lobes (we note, however, that the spectrum of this object is yet
dominated by the bright nucleus). In this case, therefore, part of the
emission filling in the absorption profile in Fig.\,\ref{f-gaussf3c}
may be locally produced in the lobes. Some contribution by nebular
\hal\ emission to the efA profile cannot be ruled out either for IRAS\,04296$+$3429,
which is known to have a $\sim$1\farcs5 extended nebula
\citep{uet00} that is unresolved in our long-slit spectrum. For objects with optically thick equatorial regions,
the nuclear region is blocked from direct view.
This is the case of IRAS\,17317$-$2743, IRAS\,17441$-$2411 and
IRAS\,20028$+$3910. Our 2D spectra of IRAS\,17441$-$2411 and
IRAS\,17317$-$2743 show, in fact, continuum emission from two extended
lobes separated by a dark equatorial region but there is no \hal\
emission locally produced in the lobes. The similar spatial
distribution along the lobes of the stellar continuum and the
structured efA \hal\ profile indicate that both the continuum and the
\hal\ line originate at the nebular core and are scattered by dust in
the lobes. This is most likely the case of IRAS\,20028$+$3910, for
which our ESI spectrum shows unresolved continuum emission arising in
the brightest of its two reflection lobes.  



As shown in Section\,\ref{proffit}, the emission filling in the
absorption in efA sources is well reproduced by a Gaussian function
with an average width of FWHM$\sim$65\kms\ (in deriving this value we
have ignored objects for which the width of the emission component is
uncertain; see Table\,\ref{t_gaussfit}).  After deconvolution with the spectral resolution in our
data (37\kms, \S\,\ref{obs}) and thermal line broadening ($\sim$15\kms\ for a
kinetic temperature of 10$^4$\,K), this value implies moderate speed
($\sim$50\kms) motions in the \hal-emitting region.
This value is larger than the typical
expansion velocity of the slow (5-15\kms) AGB wind but smaller than
the velocity of the fast pAGB outflows observed in pcyg sources
(typically $\ga$100\kms; see \S\,\ref{dis_pcyg}).  For the majority of
the objects with a reliable estimate of the Doppler shift
($\Delta\lambda$) for the emission feature (\S\,\ref{proffit}),
the emission is displaced bluewards from the narrow absorption core:
the blue-shift found ranges from $-$10 to $-$43\kms\
(Table\,\ref{t_gaussfit}). Only in one case, IRAS\,20136$+$1309, the
emission and the narrow absorption components are centered around the
same wavelength within the expected errors in our measurements.
Among the three objects with the weakest \hal\ emission features and,
therefore, most uncertain emission line parameters, two show
blue-shifted emission and one, IRAS\,04296$+$3429, red-shifted (by
$+$7\kms) emission. We note, however, that for the latter, the
parameters derived from the Gaussian analysis of the
\hal\ line are somewhat uncertain due to the weakness of the emission
component and the fact that some contamination of the efA profile by 
nebular \hal\ emission cannot be ruled out. 


We interpret the weak emission partially filling the efA \hal\ profile, which
arises in the stellar vicinity, as an indication of
present-day (i.e$.$\ pAGB) mass-loss. 
(The material in the extended nebulosities around most objects 
resulted predominantly from the large-scale mass-loss process during the AGB --
\S\,\ref{intro}.)  For some of our efA sources, the presence of
nebular material near the star is confirmed by detection of a small
number of weak non-Hydrogen emission features, e.g.\
IRAS\,04296$+$3429, IRAS\,19114$+$0002, and IRAS\,20136$+$1309 (and,
tentatively, IRAS\,19475$+$3119; \S\,\ref{res_nohi}). 
Since the stellar spectral types of the objects with efA
\hal-profiles are F and G (i.e$.$\ their central stars are relatively
cool, with temperatures ranging between 5000-7000\,K), the emission
component is likely to be formed in the de-excitation region behind a
shock wave, presumably produced by wind interaction, rather than in a
photoionized \ion{H}{2} region. The fact that the emission is
systematically doppler shifted bluewards with respect to the narrow
absorption core can be explained if there is significant occultation
of the receding part of the outflowing stellar wind.  Note that for
stellar occultation effects to be significant the size of the
\hal-emitting region has to be comparable to the radius of the star.

Alternatively, the weak \hal\ emission observed in efA sources could
arise, in principle, in a long-lived reservoir of gas near the star,
such as a rotating disk, rather than in a stellar wind. We believe,
however, that the systematic blue-shift of the emission would be more
difficult to explain under this scenario since a long-lived
circumstellar structure would necessarily have to be rotating (to
persist throughout a substantial amount of time) and, in this case,
there is no obvious reason why the emission from the receding edge of
such structure would be preferentially depressed/reduced.

Finally, although, both the narrow and broad absorption components of
the efA and pA \hal\ profiles have most likely a photospheric origin,
they are expected to arise in different layers of the stellar
atmosphere, the broad feature probably arising in deeper, denser
regions.
The observed Doppler shifts between both absorption components,
therefore, could result from the complex kinematics across the
stellar photosphere. Complex, large amplitude atmospheric motions,
involving e.g.\ the presence of shock waves propagating across the
atmosphere, are derived from asymmetrical absorption lines and radial
velocity measurements in supergiants, including those that are pPN
candidates \citep[e.g.][]{leb91,leb92}.
Such complex motions are also evidenced by variable Doppler shifts of
the Stark photospheric absorption relative to \vsys\ (see references above).


\section{Differences between pcyg and efA sources}
\label{evol}


Both pcyg and efA \hal\ profiles are interpreted in terms of on-going
pAGB mass loss.  This stresses the fact that mass-loss does not stop
after the AGB and that theoretical evolutionary models need to include
pAGB mass loss in their models.  The main difference between the \hal\
emission in pcyg and efA sources is that the strength of the line is
much larger in the former (Table\,\ref{t_pcyg}
\& \ref{t_gaussfit}). Interpreting the strength of \hal\ in terms of
mass-loss rate is extremely difficult because the former depends
critically on several parameters that are unknown, such as the
relative contribution by UV photons and shocks to the ionization of
the \hal-emitting region, whether such a region is radiation or matter
bounded, and the density distribution,
geometry, and velocity field of the gas in the stellar vicinity.  Also,
accurate modeling of the stellar atmosphere would be needed to
characterize and substract the contribution by the underlying
absorption profile to the observed
\hal\ line shape -- this is particularly significant for objects with
A- and B-type central stars. A simple, perhaps naive, interpretation is
that the more intense
\hal\ emission in pcyg sources implies larger pAGB mass-loss
rates. 
A larger amount of circumstellar material resulting from larger pAGB
mass-loss rates in pcyg sources would naturally explain their NIR
colors, which independently point to the presence of warm dust in the
vicinity of the star most likely produced by substantial present-day
mass loss.

In principle, one could think that larger pAGB mass-loss rates would
imply larger progenitor masses and, therefore, larger luminosities in pcyg
sources. The location of our targets in the HR diagram
(Fig.\,\ref{hr}) does not show particularly high values of the
luminosity for pcyg objects.
The similar distribution of pcyg and efA targets in the Galaxy does
not reveal either a significant difference in the mass of their
progenitor, although a larger sample covering a broader range in
galactic latitude would be needed to attain definitive conclusions. If
confirmed, a lack of mass/luminosity segregation between the two types
of sources would indicate that the unknown mechanism that governs
mass-loss during the pAGB phase is not strongly dependent on intrinsic
stellar characteristics but rather it may depend on extrinsic
properties like, for example, the presence of a binary
companion. 


Another difference between pcyg and efA sources concerns to the
velocity of their current stellar winds (\S\,\ref{dis_efA} and
\ref{dis_pcyg}). For efA, the weak \hal\ emission profile indicates
moderate speed motions in the nuclear
\hal-emitting region, with \vexp$\sim$50\kms. 
For pcyg sources, the values of the FWHM of the \hal\ emission core
implies larger expansion velocities in the stellar vicinity, typically
\vexp$\sim$120\kms.  Most importantly, the Doppler shift of the
absorption in the P Cygni profile suggests very fast ejections
reaching velocities of up to few$\times$10$^2$-10$^3$\kms.  This is
consistent with the earlier spectral types of pcyg sources since the
escape velocity (and, hence, the stellar wind velocity) is expected to
increase as the star evolves from the AGB to the PN phase due to the
shrinking stellar radius. The faster winds and earlier spectral types
of pcyg sources could either represent a difference in age or in the
speed traversing the pAGB path with respect to efA objects: pcyg
sources seem to be older or to have evolved faster than efA. A more
rapid evolution could again suggest more massive progenitor for pcyg
objects, but, as mentioned above, no signs of mass variance are found
in our sample. A faster evolution could also result from larger pAGB
mass loss rates in pcyg sources, which would be consistent their
strong \hal\ emission.

With regard to the extended optical nebulosities, which are detected
for most sources in our sample (Table\,\ref{t_parms}), both classes of
objects show many signs that the jet-sculpting process responsible for
the shaping of the AGB envelope to its current aspherical morphology
is active or has been active in the past. However, we have not found
any indications of pcyg sources being older (nebula-wise) than efA
targets since there are no obvious differences in their nebular
morphologies (e.g., in the diversity of shapes and/or large- and
small-scale structural components, direct visibility of the central
star, or angular size). Nevertheless, for a proper evaluation of
nebular aging, reliable estimates of the distance as well as accurate
characterization of the nebular velocity field are needed. Finally, we
would like to note that the on-going pAGB ejections probed by the
nuclear \hal\ emission in pcyg and efA sources are probably not the
same that shaped and accelerated the much more extended nebular
lobes. We don't know whether pAGB winds (similar or not to the current
ones) have been ejected in a continuous or episodic manner since the
shaping of the AGB envelope began. The fact that pcyg and efA objects
with central stars of similar late types (F and G) exhibit \hal\
emission profiles that differ both in intensity and shape may suggest
episodic pAGB ejections: pcyg would represent targets that are caught
at times of intensive jet activity leading to energetic shocks and
intense \hal\ emission, whereas efA could be currently in a status of
``mild'' wind interaction (resulted from smaller velocities and/or
densities in the pAGB wind?). Variable pAGB stellar winds are indeed
evidenced by temporal changes in the intensity and profile of some
nuclear emission lines, including \hal, in certain objects in our
sample, namely, IRAS\,19475$+$3119, IRAS\,19114$+$0002, and
IRAS\,20462$+$3416 (see \S\,\ref{var_hal} and references therein). In
pA targets, the pAGB winds that presumably shaped their aspherical AGB
envelopes may have temporarily or permanently ended.

Apart from differences in the on-going pAGB ejections, the divergence
of pcyg+pE and efA+pA sources in the IRAS color-color diagram suggests
dissimilarities in the properties of the detached dust envelope
produced by the large-scale mass-loss process during the AGB.
The separation of both types of \hal\ profile sources in the IRAS color-color
diagram seems to indicate a larger range of dust temperatures for pcyg objects. 
The different range in dust temperatures may result from: 1) a different
distribution of the grain density in the AGB CSE; 2) a different spatial
distribution of grain sizes; 3) a different grain composition, which
will result in a different power-law dependence of the
absorption/emissivity efficiency with wavelength; or any combination
of these situations.
For example, large temperature gradients are expected for optically
thick envelopes since the outside grains will remain cooler than for
optically thin CSEs due to the fact that optical-UV radiation is
heavily attenuated and therefore unable to heat up the outside
grains. Envelopes with an asymmetric dust distribution with much
larger optical depth along one preferred direction (e.g.\,along the
equatorial plane) could also result in a broader distribution of dust
temperatures since grains in regions strongly shielded from the star
radiation will remain cooler than those that are more exposed. 
Detailed modeling of the SED, including mm- and submm-wavelength data
as well as IR spectroscopy, would definitively help in improving the
characterization of the dust envelope (including its chemistry) and
mass-loss history of these objects throughout the AGB and pAGB phases.

\section{Summary}
\label{sum}

We present echelle long-slit optical spectra of a sample of evolved
intermediate-mass stars in different evolutionary stages: 5 AGB stars,
17 pPNs, and 6 yPNs. Our sample also includes the object
IRAS\,19114$+$0002, which has a controversial classification as a pPN
or a yellow hypergiant, 
and one YSO, IRAS\,05506$+$2414, which was serendipitously discovered
in our multi-wavelength survey of pPNs. (The spectrum of the latter is
presented here for completeness but it is not discussed except for
spectral typing of its central star.)  We have analyzed extracted 1D
spectra of our targets with special focus on the characteristics of
the \hal\ line profile arising in the vicinity of the central source,
i.e.\,the nebular nucleus. In this section, we summarize the main
results obtained from this work:

\begin{itemize}
  
\item[-] Fifteen objects in our sample show relatively intense \hal\ 
  emission, whereas eleven targets show \hal\ mainly in absorption.
  We have also found three sources (the AGB stars IRC$+$10011, V656
  Cas, and IK\,Tau)
  with neither \hal\ emission or absorption, and one object (the pPN
  IRAS\,19292$+$1806) in which the low S/N of the spectrum prevents us
  from determining whether \hal\ emission or absorption is present.

\item[-] Based on the shape of the \hal\ line, we have defined four
 main types of sources. Among the \hal\ emitters, those with a
 symmetric \hal\ profile are referred to as pure emission sources
 (pE), whereas objects with asymmetric P Cygni like profiles are
 denoted as pcyg. Objects with a pure absorption \hal\ profile are
 named pure absorption targets (pA). Objects with an absorption
 profile partially filled with weak emission are referred to as
 emission filled absorption sources (efA).


\item[-] The presence of \hal\ emission from the compact nebular core
 displaying either a pE, pcyg, or an efA profile is interpreted as an
 indication of on-going (i.e.\,pAGB) mass-loss most likely in the form of a stellar
 wind. The observed \hal\ profiles have been parameterized and analyzed
 to derive information on the current stellar wind and other processes
 that may be affecting the observed line shape. Among the objects that
 have already left the AGB (i.e.\,pPNs and yPNs) there are only two pA
 sources. Except for these two objects, the rest show evidence of
 current stellar winds, which supports the idea that pAGB winds are
 generally present in pPNs and yPNs and are most likely responsible
 for the nebular shaping.

\item[-] We interpret the peculiar P Cygni like profiles observed in some 
 of our objects in a similar manner as we did for He3-1475
 \citep{san01}. In this scenario the emission and absorption form in
 two distinct nebular components. The broad line emission,
 characterized by a given $intrinsic$ profile, arises from a compact
 central source. This line emission (and stellar continuum) is
 scattered by dust in the walls of the nebular lobes.  The
 blue-shifted absorption is due to neutral or partially ionized
 outflowing gas inside the lobes absorbing the scattered photons along
 the line of sight. This results in a P Cygni like profile where the
 blue-shifted absorption is produced against the intrinsic emission
 profile.

\item[-] For pE and pcyg sources the FWHM of the intense \hal\ core emission 
 indicates gas motions with velocities in the range [50:200]\kms. The
 mean velocity of the bulk of the material producing the blue-shifted
 absorption in pcyg targets varies between $V$$\sim$50 and 500\kms,
 however, larger outflow terminal velocities of up to $v_\infty
 \approx$\,1000\kms\ are observed. It is possible that a radial and
 or latitudinal velocity gradient exists in the pAGB outflows and/or that
 distinct wind components are present. 

\item[-] Broad \hal\ emission wings, with
 widths of up to $\pm$2000\kms, are observed in most pE and pcyg
 sources. The yet unclear origin of the very extended wings is
 investigated following a similar analysis to that performed by
 \cite{arr03}.  Unlike these authors, we cannot conclude that in our
 sample Raman scattering is the main mechanism for line broadening.
 The presence of fast stellar winds, which is confirmed by P Cygni
 profiles in more than one third of our targets, is another plausible
 mechanism that could contribute to the broad \hal\ wings
 observed. These winds, however, would have to reach velocities larger
 than those derived from the P Cygni absorption features ($v_\infty$)
 to explain the broad \hal\ wings.



\item[-] The \hal\ emission filling in the absorption profile of efA
 indicates moderate speed ($\sim$50\kms) motions in the nuclear
 \hal-emitting region. For most efA sources, the emission ``hump'' is
 displaced bluewards from the narrow stellar absorption feature which
 can be explained by occultation of the receding part of an outflowing
 stellar wind by the central star.

\item[-] We have found differences in the \hal\ profiles
 of some of our targets with respect to earlier observations, namely,
 IRAS\,19114$+$0002, IRAS\,19475$+$3119, IRAS\,20462$+$3416, and the
 AGB stars IRC+10216, CIT\,6, and IK\,Tau. These variations most
 likely represent real changes on the physical properties of the nuclear 
 \hal-emitting region (e.g$.$ density, excitation, ionization
 fraction, geometry and size) presumably induced by the evolution of
 the central star and/or its current stellar wind. 
  
\item[-] We briefly discuss other Hydrogen and
  non-Hydrogen lines observed towards our targets.  We note several
  prominent C$_2$ Swan bands in the AGB stars IRC$+$10216 and CIT\,6
  and the pPN IRAS\,04296$+$3429. Some of these bands have been
  reported for the first time in this work. It is also worth
  mentioning the detection of emission by the [\ion{Ca}{2}]
  $\lambda$7291,7324\AA\ doublet in the absorption-line dominated
  spectrum of the YHG IRAS\,19114$+$0002. Observation of these lines, also
  present in the spectra of IRAS\,17516$-$2525, M\,1-92, Hen\,3-1475,
  IRAS\,22036$+$5306, IRAS\,08005$-$2356, and (tentatively)
  IRAS\,19520$+$2759, is consistent with dust grain destruction by
  moderate-velocity shocks in the stellar wind. 

\item[-] We have estimated the spectral type of the central stars of the
 objects in our sample by comparing their normalized spectra with
 those of template stars from published stellar libraries to
 investigate the correlation of this fundamental stellar parameter
 with the type of \hal\ profile. In a number of objects with F- and
 G-type central stars, we found discrepancies (typically of one type)
 with respect to previous spectral type assignations obtained from low
 spectral resolution studies. Such differences may reflect the
 limitations of low resolution spectroscopy for accurate spectral
 typing, although a real time evolution of the stellar effective
 temperature cannot be ruled out.


\item[-] The stellar luminosity has been estimated from the
  luminosity-denpendent \ion{O}{1}\,7773\AA\ infrared triplet, which
  is observed in absorption in the spectrum of some of our targets (13
  out of 29). The obtained values, $L_{\rm
  bol}$$\approx$5$\times$10$^3$-10$^5$\ls, are consistent with pAGB stars
  with initial masses in the range 1-8\msun, with a significant
  fraction (10 out of 13) of objects with masses $>$3\,\msun. Such
  massive pAGB objects represent $\sim$30\% of our whole sample. The
  rest of the sources, which is also the majority ($\sim$70\%), may
  well have low-mass ($\la$3\msun) progenitors -- this may be part of
  the reason why the \ion{O}{1} triplet is not observed in
  absorption. For the highest masses/luminosities derived in our
  sample, the theory of pAGB stellar evolution predicts rapid changes
  of the effective stellar temperature that should be observable in a
  time-scale of $\approx$10-100\,years.

\item[-] We investigated correlations between the type of \hal\ 
  profile and some stellar and envelope parameters. The shape of
  the \hal\ line emission is correlated with the stellar spectral type as
  well as the the NIR and IRAS colors.
  No correlation has been found with the chemistry, galactic latitude,
  or nebular morphology.

\item[-] All sources in which \hal\ is seen mainly in absorption 
 (i.e., pA and efA) have F-G type central stars, whereas sources with
 intense \hal\ emission (i.e., pE and pcyg) span a larger range of
 spectral types from O to G, with a relative maximum around B, and
 also including very late C types. The measured equivalent widths of
 the \hal\ emission in objects with O- and B-type stars are consistent
 with UV stellar radiation being the main ionizing agent. \hal\
 emitters with cooler central stars lack enough ionizing radiation,
 therefore, the emission component in these cases is likely to be
 formed in the de-excitation region behind a shock wave, presumably
 produced by wind interaction.

\item[-] Pcyg and pE sources are found to exhibit a larger $J-K$ 
 color excess than pA and efA objects. Moreover, while the NIR colors
 of efA+pA sources are consistent with the major contribution to the
 emission in the NIR bands being the reddened stellar photosphere, the
 position of pcyg+pE sources in the ($J-H$)-($H-K$) color-color
 diagram highlights the presence of warm dust. This component of warm
 dust probably results from substantial present-day mass-loss evidenced by
 the strong \hal\ emission of pE and pcyg targets.
  
\item[-] Intense \hal\ emitters (i.e.\,pcyg+pE) and objects with \hal\ mainly in absorption (i.e.\,
 pA+efA) also segregate in the IRAS color-color diagram in a way that
 the former have dust grains with a larger range of temperatures.
 Such a different temperature range may result from: 1) a different
 distribution of the grain density in the AGB envelope; 2) a different
 spatial distribution of grain sizes; 3) a different grain
 composition; or any combination of these situations.

\item[-] The differences found between pE+pcyg and pA+efA sources 
 described above indicate dissimilarities in their pAGB and AGB
 mass-loss histories. The intense \hal\ emission and NIR color excess
 of pE and pcyg sources may indicate larger pAGB mass-loss rates
 compared with those in efA and pA targets.  The lack of a
 mass/luminosity segregation of the two profile sources suggests that
 pAGB mass-loss is not very dependent on intrinsic stellar properties
 but may be dictated by extrinsic factors like, for example, the
 presence of a binary companion.
 The faster winds and earlier spectral types of the central stars of
 pcyg and pE sources suggest that these are older and/or have a larger
 speed traversing the pAGB evolutionary path than pA and efA.

\item[-] We have not found any indications of the extended optical
 nebulosities (observed in most objects) being older for pcyg+pE
 sources than for efA+pA targets. 
 Such nebulosities show many signs that the jet-sculpting process of the AGB
 CSE is currently active or has been active in the past in all cases.
 For the only two pA targets in our sample, IRAS\,19477$+$2401 and
 IRAS\,17150$-$3224, the winds that shaped their AGB CSEs to its
 current aspherical morphology are not seen at present. The on-going
 pAGB winds probed by the nuclear \hal\ emission in pE+pcyg and efA
 objects, which represent the vast majority of our sample, may not be
 the same that carved and accelerated the much more extended (probably
 older) optical lobes, however, shaping of the innermost layers of the
 AGB envelope (by interaction with the present-day pAGB winds) is
 probably still at work.

 \item[-] In principle, pAGB winds may have been ejected in a
 continuous or episodic manner since the shaping of the AGB envelope
 began. The fact that pcyg and efA objects with central stars of
 similar late types (F and G) exhibit quite different \hal\ emission
 profiles may suggest episodic pAGB ejections: pcyg would represent
 targets that are caught at times of intensive jet activity leading to
 energetic shocks and intense \hal\ emission, whereas efA could be
 currently in a status of ``mild'' wind interaction (resulted from
 smaller velocities and/or densities in the pAGB wind?). In pA
 targets, the pAGB winds that presumably shape their aspherical AGB envelopes 
 may have temporarily or permanently ended.

\end{itemize}



\acknowledgments

Most of data presented herein were obtained at the W.M. Keck
Observatory, which is operated as a scientific partnership among the
California Institute of Technology, the University of California, and
NASA. The Observatory was made possible by the generous financial
support of the W.M. Keck Foundation. The authors wish to recognize and
acknowledge the very significant cultural role and reverence that the
summit of Mauna Kea has always had within the indigenous Hawaiian
community.  We are most fortunate to have the opportunity to conduct
observations from this mountain.  We are very grateful to B.F$.$
Madore for kindly providing part of his time at the 6.5-m Magellan-I
(Baade) telescope in Las Campanas Observatory to this project.  The
Magellan telescope is operated by a consortium consisting of the
Carnegie Institution of Washington, Harvard University, MIT, the
University of Michigan, and the University of Arizona. We also thank
Gregory Walth and Daniel Kelson for processing the MIKE spectra for
us. This work has been partially performed at the California Institute
of Technology and the Department of Molecular and Infrared
Astrophysics of the {\sl Instituto de Estructura de la Materia, CSIC,}
and has been partially supported by Caltech optical observatories and
by the Spanish MCyT under project AYA\,2006-14876 and the Spanish MEC
under project PIE 200750I028.  RS thanks NASA for partially funding
this work through an LTSA award (no. 399-20-40-06) and ADP award
(no. 399-30-00-08); RS also received partial support for this work
from HST/GO awards (nos. GO-09463.01, 09801.01, and 10185.01) from the
Space Telescope Science Institute (operated by the Association of
Universities for Research in Astronomy, under NASA contract
NAS5-26555). RS's contribution to the research described in this
publication was carried out at the Jet Propulsion Laboratory,
California Institute of Technology, under a contract with NASA.  AGdP
is partially financed by the MAGPOP EU Marie Curie Research Training
Network and by the Spanish {\sl Programa Nacional de Astronom\'{\i}a y
Astrof\'{\i}sica} under grants AYA2003-01676 and AYA2006-02358.  We
acknowledge the use of data from the UVES Paranal Observatory Project
(ESO DDT Program ID 266.D-5655).  This research has made use of the
SIMBAD database, operated at CDS, Strasbourg, France, and NASA's
Astrophysics Data System.



{\it Facilities:} \facility{Keck:II (ESI), Magellan:Baade (MIKE)}



\appendix 
\section{Discrepant stellar spectral type classification}
\label{starevol}

The stellar spectral type estimated from our ESI data is different
from previous assignations for several objects
(\S\,\ref{specclass}). In the majority of the cases, we find earlier
spectral types (i.e$.$\,larger \teff) than those reported in the past.
The most extreme case is IRAS\,19306$+$1407, which has been classified
as G5, but displays clear signatures of a B-type star in our ESI data
(see below).
There is also a total of four pPNs with central stars formerly
classified as G but with ESI spectra most consistent with F types. In
two cases the spectral class derived from this work is later than in
previous measurements, namely, IRAS\,17317$-$2743 and
IRAS\,19477$+$2410, which have gone from a past mid-F to a current G4
and G0 classification, respectively. Incidentally, we note that these
two objects are efA+pA, whereas sources for which our spectral type is
earlier than in former works are all pcyg+pE.

Discrepancies between spectral types obtained from low- and
high-resolution studies has been noticed for other pAGB objects and is
discussed by, e.g$.$, \cite{dec98} and \cite{rh99}.  Among the objects
with discrepant past and our current spectral type classifications,
IRAS\,04296$+$3429 is the only one with additional high-resolution
spectroscopic line studies in the literature. The fact that the
spectral type and \teff\ for IRAS\,04296$+$3429 determined by such
studies (including this work) are all in very good agreement but
systematically differ from the spectral classification from the low
resolution spectrum indicates that the changes observed do not
correspond to a real time evolution of the stellar \teff, but rather
reflect the inadequacy/limitations of low-resolution spectra for
accurate spectral-type determinations. (Note that the previous
high-resolution data of this object were taken at different epochs,
from 1988 to 2003, and the low-resolution spectrum was obtained in
1991.) This could also well be the case for objects with low and
high-resolution spectral types in disagreement by one type,
e.g$.$ objects previously classified as F (G) versus our current
classification as G (F). In other words, we assume that the typical
error of the spectral type derived from low-resolution spectra can be
of one type but unlikely larger than that. (Note, however, that a real
time variation in \teff\ cannot be completely ruled out.)


According to this, the spectral type change observed in
IRAS\,19306$+$1407 can hardly be accounted for by the different
spectral resolution. Although it is not impossible, we believe that
the different spectral type does not result from the rapid evolution
of the central star towards higher temperatures. There are remarkable
differences between our spectrum and that observed by \cite{sua06}
that suggest that the spectrum reported by these authors probably did
not correspond to IRAS\,19306$+$1407.
This is mainly supported by the absence of the deep \ion{Na}{1}\,D
doublet lines at $\sim$5900\AA\ in the spectrum by \cite{sua06}. The
radial velocity derived from the sodium lines in our data,
\vlsr$\sim$30\kms, is very different from that measured in stellar
absorption lines, such as \ion{C}{2}\,6578,6583\AA\ and
\ion{He}{1}\,6678\AA, which yield \vlsr$\sim$90-100\kms, in agreement
with the systemic velocity of the source derived from CO measurements
(Table\,\ref{t_pcyg}). The velocity of the sodium lines indicates that
their origin is most likely interstellar and, therefore, their
absence in Suarez et al.'s spectrum, which is not explained by their
lower S/N, suggests that the object observed by these authors is not
IRAS\,19306$+$1407. (We note that if the \ion{Na}{1}\,D lines in our
spectrum were stellar then they should have been even deeper in a G5
star.) Another indication of the possible object misidentification is
the absence in Suarez et al.'s data of the broad diffuse bands at
4430, 5780, 5797, 6284, 6614, 6993 and 7224\AA\ observed by us, which
may also be interstellar.

In spite of the above reasoning, we would like to note that it is not totally
unrealistic to expect spectral changes due to rapid pAGB evolution of
the central star of IRAS\,19306$+$1407 in a time scale of
$\approx$10\,yr. In fact, the high luminosity derived from the deep
\ion{O}{1} infrared triplet indicates a final mass of the central
star in the range [$>$0.84-0.94\msun], which for the
standard/empirical initial-final mass relationship, implies an initial
mass of $M_{\rm ZAMS}$$\sim$5-8\msun.  According to the theoretical
evolutionary models for massive pAGB stars by \cite{blo95}, a
5-8\msun\ star reaches \teff$\sim$20,000\,K in only 80-10\,years,
respectively. (For comparison, traversing the same spectral interval
would take $\sim$10$^3$\,yr for a $M_{\rm ZAMS}$=2\,\msun.)

To the best of our knowledge, evolution towards higher temperatures
has been previously reported for two pAGB stars: SAO 85766
(IRAS\,18062$+$2410) and SAO 244567 (Hen\,3-1357) \citep[][and
references therein]{par00}. For SAO 85766, an increase of the
temperature from $\sim$8500\,K to $\sim$22000\,K in less than 25 yr is
found, and SAO 244567 has turned into a young PN within the last 20-30
years and its central star appears to be evolving rapidly into a white
dwarf.  A very fast evolution of the temperature of the central star
of the YHG IRC$+$10420 has been also reported by \cite{klo97} who
measured an increase of \teff\ from $\sim$5700 to $\sim$8600\,K in
only 4 years during 1992 to 1996.

\clearpage
\begin{table}
\small
\begin{center}
\caption{Journal of observations\label{t_log}}
\begin{tabular}{lccccrr}
\tableline\tableline
IRAS & Other & R.A.\ (2000) & Dec.\ (2000) & Date observed\ & Exposure
& P.A.\ \\ name & name & (h m s) & ($^{\rm o}$ ' '') &
($yy$-$mm$-$dd$) & time (s) & (\degr) \\
\tableline
01037$+$1219 & IRC$+$10011      & 01:06:25.98 & $+$12:35:53.1 & 2004$-$11$-$08 & 1800  & 135  \\
02316$+$6455 & V656 Cas         & 02:35:44.67 & $+$65:08:58.7 & 2004$-$11$-$08 & 700   & 166  \\
03507$+$1115 & IK Tau           & 03:53:28.87 & $+$11:24:21.7 & 2004$-$11$-$08 & 1250  & 120  \\    
04296$+$3429 &                  & 04:32:56.97 & $+$34:36:12.4 & 2004$-$11$-$08 & 1500  & 95   \\
05506$+$2414-$Sa$ &                     & 05:53:43.56 & $+$24:14:44.7 & 2004$-$11$-$08 & 1800  & 136  \\
05506$+$2414-$Sb$ &                     & 05:53:43.18 & $+$24:14:44.1 & 2004$-$11$-$08 & 1200  & 45  \\
08005$-$2356 &                  & 08:02:40.71 & $-$24:04:42.7 & 2004$-$11$-$08 & 1320  & 132  \\
09452$+$1330 & IRC$+$10216, CW Leo & 09:47:57.41 & $+$13:16:43.6 & 2004$-$11$-$08 & 3600  & 22   \\
10131$+$3049 & CIT$-$6, RW LMi  & 10:16:02.29 & $+$30:34:19.1 & 2004$-$11$-$08 & 1470  & 20   \\ 
17150$-$3224 & AFGL 6815        & 17:18:19.86 & $-$32:27:21.6 & 2003$-$04$-$27 & 1200 & 94\tablenotemark{\dagger}\\ 
17317$-$2743 &                  & 17:34:53.29 & $-$27:45:11.5 & 2003$-$06$-$03 & 900  & 7 \\
17423$-$1755 & Hen 3$-$1475     & 17:45:14.19 & $-$17:56:46.9 & 2003$-$06$-$04 & 600  & 135 \\  
17440$-$3310 &                  & 17:47:22.72 & $-$33:11:09.3 & 2003$-$04$-$28 & 3600 & 115\tablenotemark{\dagger\dagger}\\       
17441$-$2411 & Silkworm nebula  & 17:47:13.49 & $-$24:12:51.4 & 2003$-$06$-$03 & 2700 & 15 \\
17516$-$2525 &                  & 17:54:43.35 & $-$25:26:28.0 & 2003$-$06$-$02 &  1920 & 164  \\
18167$-$1209 &                  & 18:19:35.50 & $-$12:08:08.2 & 2003$-$06$-$04 & 800  & 3   \\  
19024$+$0044 &                  & 19:05:02.06 & $+$00:48:50.9 & 2003$-$06$-$02 &  2400 & 138  \\
19114$+$0002 & AFGL\,2343       & 19:13:58.61 & $+$00:07:31.9 & 2003$-$06$-$02 &  26   &  55  \\
19292$+$1806 &                  & 19:31:25.37 & $+$18:13:10.3 & 2004$-$11$-$08 & 600   & 137.4\\
19306$+$1407 &                  & 19:32:55.09 & $+$14:13:36.9 & 2003$-$06$-$04 & 1800 & 11  \\  
19343$+$2926 & M1$-$92          & 19:36:18.91 & $+$29:32:50.0 & 2003$-$06$-$02 &  1800 & 131  \\
19374$+$2359 &                  & 19:39:35.55 & $+$24:06:27.1 & 2003$-$06$-$04 & 1800 & 15  \\  
19475$+$3119 &                  & 19:49:29.56 & $+$31:27:16.3 & 2003$-$06$-$02 &  40   & 112 \\
19477$+$2401 & Cloverleaf nebula& 19:49:54.91 & $+$24:08:53.3 & 2003$-$06$-$03 & 1800 & 91 \\
19520$+$2759 &                  & 19:54:05.87 & $+$28:07:40.6 & 2003$-$06$-$04 & 1500 & 139 \\
20028$+$3910 &                  & 20:04:35.98 & $+$39:18:44.5 & 2004$-$11$-$08 & 600   & 165  \\ 
20136$+$1309 &                  & 20:16:00.51 & $+$13:18:56.3 & 2003$-$06$-$02 &  1500 &  20   \\
20462$+$3416 & LS\,II +34 26    & 20:48:16.64 & $+$34:27:24.3 & 2003$-$06$-$02 & 1660 &  63   \\
21282$+$5050 &                  & 21:29:58.48 & $+$51:04:00.3 & 2003$-$06$-$04 & 600  & 165 \\
22036$+$5306 &                  & 22:05:30.28 & $+$53:21:32.8 & 2003$-$06$-$04 & 2100 & 66  \\
22574$+$6609 &                  & 22:59:18.36 & $+$66:25:48.3 & 2004$-$11$-$08 & 3900  & 130  \\

\tableline
\end{tabular}
\tablenotetext{\dagger}{PA changed from 85.9 to 102.1\degr}
\tablenotetext{\dagger\dagger}{PA changed from 110.3 to 121.5\degr}
\tablecomments{All sources were observed with Keck-II/ESI except for IRAS\,17150$-$3224 and 
IRAS\,17440$-$3310, which were observed with Magellan-I/MIKE (see \S\,\ref{obs}); Coordinates are 2MASS J2000.}
\end{center}
\end{table}

\clearpage
\begin{table}
\tiny
\begin{center}
\caption{Stellar and envelope parameters for the different types of H$\alpha$ profiles\label{t_parms}}
\begin{tabular}{lcccrlcl}
\tableline\tableline
IRAS    &  Object & Spectral     & f12/f25 & f60  & Morphology\tablenotemark{1} &  C/O\tablenotemark{2} & References \\
name    &  Class  & Type         &         & (Jy) & (optical/NIR)               &                       &        \\
\tableline
\multicolumn{3}{c}{\it Pure emission profiles (pE)} &&&& \\
09452$+$1330  & AGB & C9.5                  &2.06 &5652.0 &Bw  & C    & 24 \\ 
10131$+$3049\tablenotemark{*} & AGB & C4,3e &2.72 &273.6  &B$\star$   & C  &  7, 25 \\ 
19374$+$2359  & yPN & {\bf B3-6\,I}              &0.24 &70.9   &B  & O  & 15, 24, 23 \\ 
\multicolumn{3}{c}{\it P Cygni profiles (pcyg)}  &&&&\\
05506$+$2414-$Sa$  & YSO\tablenotemark{\dag}   & {\bf G9-K2} &0.23 &103.4  &I & O  &  15, 16 \\ 
08005$-$2356  & pPN & F5Ie                      &0.35 &29.8   &B$\star$    & C/O &  23, 24 \\ 
17423$-$1755  & yPN & Be                        &0.25 &63.7   &Bw$\star$    & O  &  15, 23 \\ 
17516$-$2525  & pPN & {\bf O-B?}                 &0.45 &100.1  &S       & O  &  17, 26 \\ 
19024$+$0044  & pPN & G0-5                      &0.06 &42.5   &Mw     & O  &  14, 15 \\ 
19306$+$1407  & yPN & {\bf B0-1\,I}                     &0.06 &31.8   &B$\star$      & C/O & 8, 15 \\ 
19343$+$2926  & pPN & B2 (+F5)                          &0.29 &118.0  &Bw$\star$       & O & 1, 15 \\ 
19520$+$2759  & yPN & {\bf O9\,I?}              &0.39 &207.0  &S      & O  & 4, 17 \\ 
20462$+$3416  & yPN & B1Iae                     &0.02 &12.1   &E$\star$      & O  & 4, 22, 24 \\ 
21282$+$5050  & yPN & O9.5,WC11                 &0.68 &33.4   &M$\star$      & C  & 2, 15 \\ 
22036$+$5306  & pPN & F4-7                      &0.18 &107.2  &Bw$\star$  & O  & 13, 15 \\ 
22574$+$6609  & pPN & {\bf A1-6\,I}            &0.31 &20.6   &I/Bw  &C  & 18, 21 \\ 
\multicolumn{3}{c}{\it Emission filled absorption profiles (efA)}   &&&&\\
04296$+$3429\tablenotemark{**} & pPN & {\bf F3\,I} &0.28 &15.4   &Bw($\star$?)       & C  &  5, 11, 15, 23, 24 \\ 
17317$-$2743  & pPN & {\bf G4\,I}                       &0.04 &29.5   &Bw\tablenotemark{\dag\dag}  & O  & 10, 20, 22 \\ 
17440$-$3310  & pPN & {\bf F3\,I}                 &0.20 &30.1   &B$\star$(w?)    & O  & 15, 20 \\ 
17441$-$2411\tablenotemark{**} & pPN & F4-5I     &0.22 &106.0  &Bw$\star$   & C? & 11, 15, 22, 23, 24 \\ 
18167$-$1209\tablenotemark{**} & pPN & {\bf F7\,I} &0.16 &21.3    & S     & O  & 10, 17 \\ 
19114$+$0002  & pPN/YHG? & F5-7Ia               &0.05 &516.0   &E$\star$  & O  & 12, 22, 24, 27 \\ 
19475$+$3119  & pPN & F3Ib                      &0.01 &55.8    &M$\star$  & O  & 15, 23 \\ 
20028$+$3910  & pPN & {\bf F3-7\,I}                     &0.20 &143.0   &Bw    & C  &  15, 23, 24 \\ 
20136$+$1309  & pPN & {\bf F3-7\,I}                     &0.44 &2.1     &S    & O? &  21, 23, 24 \\ 
\multicolumn{3}{c}{\it Pure absorption profiles (pA)}  &&&&\\
17150$-$3224  & pPN & {\bf F3-7\,I}                    &0.18 &268.3  &Bw$\star$    & O  & 3, 15, 24 \\ 
19477$+$2401  & pPN &   {\bf G0\,I}       &0.20 &27.1   &M   & C  & 15, 21, 22 \\ 
\multicolumn{3}{c}{\it \hal\ non-detections}  &&&&\\
01037$+$1219  & AGB & M7               &1.19 &215.2  &E   & O  & 7, 9, 15 \\ 
02316$+$6455  & AGB & M8               &1.53 &45.8   &S   & O  &  7, 17 \\ 
03507$+$1115  & AGB & M6e-M10e         &1.95 &332.1  &I   & O  &  7, 9 \\ 
19292$+$1806  & pPN &{\bf B?}          &0.10 &28.8:  &Bw  & O & 15, 18 \\ 
\tableline
\end{tabular}
\tablenotetext{1}{Main morphology descriptors defined by \cite{sahs07}. B\,=\,bipolar, M\,=\,multipolar, E\,=\,elongated, I\,=\,irregular, S\,=\,stellar (i.e$.$\ unresolved in $HST$
images), w\,=\,central obscuring waist, $\star$\,=\,central star evident.}
\tablenotetext{2}{O\,=\,Oxygen rich; C\,=\,Carbon rich; C/O\,=\,mixed chemistry.}
\tablenotetext{*}{Uncertain profile due to weak \hal\ emission blended with other lines.}
\tablenotetext{**}{Uncertain profile: could also be pA.}
\tablenotetext{\dag}{YSO serendipitously discovered in our survey of pPNs -- see \S\,\ref{intro}.}
\tablenotetext{\dag\dag}{No optical image available; our long-slit spectrum shows two nebulosities 
separated by a dark lane, i.e$.$, most likely a bipolar nebula.}
\tablecomments{Boldfaced spectral types have been estimated in this work (see \S\,\ref{specclass}).}
\tablerefs{
(1) \cite{arr05}; 
(2) \cite{crow98}; 
(3) \cite{dav05}; 
(4) \cite{gled05};
(5) \cite{klo99}; 
(6) \cite{klo02}; 
(7) \cite{loup93};
(8) \cite{low07}; 
(9) \cite{mau06}; 
(10) \cite{nym98}; 
(11) \cite{opp06}; 
(12) \cite{rh99}; 
(13) \cite{sah03}; 
(14) \cite{sah05}; 
(15) \cite{sahs07}; 
(16) \cite{sah08}; 
(17) Sahai et al., in prep ($HST$ GO 9463, 10185); 
(18) \cite{sanao}; 
(19) \cite{sch02}; 
(20) \cite{sev02}; 
(21) \cite{su01}; 
(22) \cite{sua06}; 
(23) \cite{szc07}: ``The Torun catalogue of Galactic post-AGB and related objects''; 
(24) \cite{uet00}; 
(25) \cite{uet07}; 
(26) \cite{vdveen89}; 
(27) \cite{zac96}.}
\end{center}
\end{table}

\clearpage
\begin{table}
\begin{center}
\caption{Parameters of the \hal\,profile for pcyg and pE sources \label{t_pcyg}}
\begin{tabular}{cc|cccr|ccc}
\tableline\tableline
&& \multicolumn{4}{c}{\underline{Emission}} & \multicolumn{3}{c}{\underline{Absorption}} \\ 
IRAS  &  $V_{\rm sys}$(LSR) &  $V_{\rm e}$(LSR) & FWHM   & FWZI   & $W_\lambda$ & $V$ & $V_{\rm max}$ & FWHM \\ 
name  &  (\kms)             & (\kms)       & (\kms) & (\kms) &  (\AA)      & (\kms)                & (\kms)                 & (\kms) \\
\tableline
\multicolumn{1}{c}{\sl pcyg srcs} &&&&&& \\
08005$-$2356  & $+$47$^{(4)}$ & $+$101 & 190 & $\sim$2400  &14  & $-$85 & $-$170  & 80 \\        
17423$-$1755  & $+$48$^{(1)}$ & $+$89  & 170 & $\sim$4000  &110 & \tablenotemark{\dag}      &\tablenotemark{\dag}       &\tablenotemark{\dag} \\ 
17516$-$2525  & $-$15$^{(6)}$ & $+$35  & 100 & $\sim$1400  &105 & $-$40   & $-$110  & 70 \\   
19024$+$0044 & $+$50$^{(2)}$  &$+$101  & 90  & $\sim$2400  &40  & $-$100  & $-$170  & 80 \\   
19306$+$1407 & $+$90$^{(5)}$  &$+$116  & 60  & $\sim$2600  &15  & $-$115  & $-$170  & 50 \\   
19343$+$2926 & $-$1$^{(1)}$   &$+$19   & 170 & $\sim$3000  &180 & $-$530  & $-$750  & 220 \\   
19520$+$2759 & $-$15$^{(5)}$  &$-$5    & 170 & $\sim$2600  &130 & $-$260  & $-$420  & 120 \\    
''           &''              &''      &''        &''      &''  & $-$530  & $-$800  & 250 \\    
20462$+$3416 &$-$75$^{(8)}$   & $-$43  & 80  & $\sim$2800  &15  & \tablenotemark{\dag}      &\tablenotemark{\dag}       &\tablenotemark{\dag} \\ 
21282$+$5050 &$+$18$^{(1)}$   & $+$17  & 50  & $\sim$500   &23  & \tablenotemark{\dag}      &\tablenotemark{\dag}       &\tablenotemark{\dag} \\ 
22036$+$5306 &$-$43$^{(3)}$   & $+$0   & 220 & $\sim$2500  &30  & $-$150  & $-$250  & 110 \\   
22574$+$6609 &$-$64$^{(1)}$   & $-$41  & 70  & $\sim$400   &3   & $-$50   & $-$110  & 50 \\   
\multicolumn{1}{c}{\sl pE srcs} &&&&&& \\
19374$+$2359 &$-$35$^{(5)}$   & $-$7   & 90  & [600:2000]  & 4 &...       &...       &... \\        
09452$+$1330 &$-$25$^{(7)}$   & $-$32  & 65  & $\sim$150   & 4 &...       &...       &... \\        
\tableline
\end{tabular}
\tablenotetext{\dag}{Absorption partially masked by emission within the PSF in our ESI spectra (see \S\,\ref{res_hi})}
\tablecomments{References for $V_{\rm sys}$(LSR): (1) Bujarrabal et al.\ 2001; (2) Sahai et al.\ 2005; (3) Sahai et al.\ 2006; (4) Slijkhuis et al.\ 1991; (5) CO data from 
S\'anchez Contreras et al., in prep.; (6) Nyman et al.\ 1998; (7) Teyssier et al.\ 2006; (8) \cite{tur84}}
\end{center}
\end{table}

\clearpage
\begin{table}
\begin{center}
\caption{Gaussian analysis of the \hal\,profile for efA and pA sources \label{t_gaussfit}}
\begin{tabular}{c|cc|ccc|ccc}
\tableline\tableline
& \multicolumn{2}{c}{\underline{Narrow absorption}} & \multicolumn{3}{c}{\underline{Broad absorption}} & \multicolumn{3}{c}{\underline{Emission}}\\
IRAS        & $W_\lambda$   & FWHM & $W_\lambda$    & $\Delta\lambda$  & FWHM & $W_\lambda$ & $\Delta\lambda$ & FWHM \\
name        & (\AA)    &(\kms)&     (\AA)     & (\kms)             &(\kms)&     (\AA)  & (\kms)            & (\kms)\\ 
\tableline
17317$-$2743  & 2.22 & 140 & 2.25 & $+$1(21)  & 880 & 0.73 & $-$28(6) &  60 \\      
17440$-$3310* & 3.94 &  70 & 0.98 & $-$10(3)  & 200 & 4.89 & $+$4(1) & 100 \\         
19114$+$0002* & 1.96 &  80 & ...  & ...       & ... & 1.56 & $+$8(3) & 110 \\         
19475$+$3119  & 0.77 &  60 & 2.22 & $-$1(3)   & 720 & 0.60 & $-$20(2) &  50 \\        
20028$+$3910  & 0.67 &  50 & 1.27 & $-$50(17) & 490 & 0.32 & $-$40(10)&  70 \\        
20136$+$1309  & 1.09 &  60 & 1.10 & $-$32(3)  & 580 & 0.79 & $-$1(2)  &  80 \\        
04296$+$3429  & 0.65 &  60 & 2.04 & $+$10(3)  & 640 & 0.21 & $+$7(5)  & 125 \\        
''            & 0.54 &  50 & 1.97 & $-$1(3)   & 710 & ...    & ...      & ... \\              
17441$-$2411  & 0.61 &  50 & 1.86 & $-$13(3)  & 730 & 0.09 & $-$34(8) &  60 \\       
''            & 0.54 &  50 & 1.88 & $-$18(3)  & 750 & ...    & ...      & ... \\              
18167$-$1209  & 0.69 &  55 & 1.94 & $-$21(2)  & 530 & 0.16 & $-$10(3) &  80 \\        
''            & 0.56 &  50 & 1.89 & $-$27(2)  & 520 & ...    & ...      & ... \\             
17150$-$3224  & 0.50 &  50 & 1.36 & $+$30(13) & 530 & ...    & ...      & ... \\              
19477$+$2401  & 1.09 & 110 & ...  & ...       & ... & ...    & ...      & ... \\              
\tableline
\end{tabular}
\tablenotetext{*}{Parameters particularly uncertain (see text \S\,\ref{proffit}).} 
\tablecomments{Values of $\Delta\lambda$ are relative to the center of the narrow absorption core.  $\Delta\lambda$ errors (in \kms) are indicated in parentheses.}
\end{center}
\end{table}

\clearpage
\begin{table}
\begin{center}
\caption{Equivalent widths of the \ion{O}{1} 7773\AA\ triplet and derived magnitudes \label{t_oi}}
\begin{tabular}{ccrcccc}
\tableline\tableline
IRAS & $W_\lambda$(\ion{O}{1}) & $M_V$ & B.C. & $L_{\rm bol}$ & $L/D^2$ & $D$ \\
name & (\AA)           & (mag) & (mag)& (\ls) & (\ls\,kpc$^{-2}$) & (kpc)  \\ 
\tableline
04296$+$3429 & 1.62 & $-$5.83                 &$-$0.00 & 1.7E+04                 &  280 & $<$8  \\
17441$-$2411 & 1.81 & $-$6.42                 &$-$0.03 & 3.0E+04                 &  810 & $<$6  \\
18167$-$1209 & 1.93 & $-$6.81                 &$-$0.09 & 4.5E+04                 &  100 & $<$21 \\
19024$+$0044 & 1.69 & $-$6.04                 &$-$0.21 & 2.5E+04                 &  250 & $<$10 \\
19114$+$0002 & 3.00 & $-$10.2                 &$-$0.03 & 9.5E+05\tablenotemark{\dag} & 4950 & $<$14\tablenotemark{\dag} \\
19306$+$1407 & 0.99 & $-$3.81                 &$-$2.14 & 1.9E+04\tablenotemark{\dag\dag}&  320 & $<$8\tablenotemark{\dag\dag}  \\
19374$+$2359 & 1.98 & $-$6.95                 &$-$0.95 & 1.1E+05\tablenotemark{\dag\dag}&  450 & $<$13\tablenotemark{\dag\dag}$^,$\tablenotemark{\ast} \\
19475$+$3119 & 2.10 & $-$7.35                 &$-$0.03 & 6.9E+04                 &  380 & $<$13 \\
19477$+$2401 & 1.20 & $-$4.47                 &$-$0.15 & 5.5E+03                 &  240 & $<$5\tablenotemark{\ast} \\ \\ 
20028$+$3910 & 2.04 & $-$7.15                 &$-$0.03 & 5.8E+04                 &  950 & $<$8 \\ 
20136$+$1309 & 1.26 & $-$4.67                 &$-$0.03 & 5.9E+03                 &   75 & $<$9 \\ 
20462$+$3416 & 0.67 & $-$2.81                 &$-$2.14 & 7.4E+03                 &  130 & $<$8 \\ 
22574$+$6609 & 2.10 & $-$7.34                 &$-$0.13 & 7.6E+04\tablenotemark{\dag\dag} &  170 & $<$21\tablenotemark{\dag\dag} \\ 
\tableline
\end{tabular}
\tablenotetext{\dag}{Uncertain, outside of the $M_V$ applicability domain (see \S\,\ref{lum}).} 
\tablenotetext{\dag\dag}{Uncertain, outside of the spectral type applicability domain (see \S\,\ref{lum}).} 
\tablenotetext{\ast}{Interstellar $A_V$ expected to be very high.}

\end{center}
\end{table}

\clearpage


\figsetstart
\figsetnum{1}
\figsettitle{Observed spectra}

\figsetend

\clearpage
\thispagestyle{empty}
\setlength{\voffset}{-25mm}
\clearpage
\setlength{\voffset}{0mm}

\end{document}